\documentclass[10pt,journal,compsoc]{IEEEtran}

\usepackage{graphics}
\usepackage{subfigure}
\usepackage{amssymb}
\usepackage{amsmath}
\usepackage{amsthm}
\usepackage{amsfonts}         
\usepackage{color}            % importation depuis Xfig
\usepackage{epic}             % compl?mente la suite
\usepackage{eepic}            % remplissage Xfig
\usepackage{rotating}         % texte dans les images
\usepackage{type1cm}          % polices de taille arbitraire
\usepackage{url}
\usepackage{epsfig}
\usepackage{amsmath}
\usepackage{graphicx}
\usepackage{graphics}
\usepackage{epsfig}
\usepackage{latexsym}
\usepackage{amsfonts}
\usepackage{paralist}
\usepackage{xspace}
\usepackage{mathrsfs}
\usepackage{psfrag}
\usepackage{color}
\usepackage[ruled]{algorithm}
\usepackage[noend]{algorithmic}

\newtheorem{definition}{Definition}
\newtheorem{theorem}{Theorem}
\newtheorem{lemma}{Lemma}

\def\participant{{\mathcal{P}}}

\def\HPK{\mathcal{PK}}
\def\HSK{\mathcal{SK}}
\def\MK{\mathcal{K}}
\def\SK{\textit{EK}}

\def\ourprotocol{\xspace{\textbf{PDA}}}

\begin{document}

\title{{\textbf{PDA}}: Semantically Secure Time-Series Data Analytics with Dynamic Subgroups}

%\title{Efficient Privacy Preserving Data Aggregation without Secure Channels}

\author{Taeho Jung$^1$, Junze Han$^1$, Xiang-Yang Li$^{12}$\thanks{*Dr. Xiang-Yang Li (xli@cs.iit.edu) is the contact author of this paper.}
\\$^1$Department of Computer Science, Illinois Institute of Technology, Chicago, IL
\\$^2$Department of Computer Science and Technology, TNLIST, Tsinghua University, Beijing
}

\IEEEtitleabstractindextext{%


\begin{abstract}
Third-party analysis on private records is becoming increasingly important due to the widespread data collection for various analysis purposes. However, the data in its original form often contains sensitive information about individuals, and its publication will severely breach their privacy. In this paper, we present a novel Privacy-preserving Data Analytics framework \ourprotocol, which allows a third-party aggregator to obliviously conduct many different types of polynomial-based analysis on private data records provided by a dynamic sub-group of users. Notably, every user needs to keep only $O(n)$ keys to join data analysis among $O(2^n)$ different groups of users, and any data analysis that is represented by polynomials is supported by our framework. Besides, a real implementation shows the performance of our framework is comparable to the peer works who present ad-hoc solutions for specific data analysis applications. Despite such nice properties of \ourprotocol, it is provably secure against a very powerful attacker (chosen-plaintext attack) even in the Dolev-Yao network model where all communication channels are insecure.
\end{abstract}
}
\maketitle

%\footnotetext[1]{The research of authors is partially supported by NSFC under Grant No. 61170216, No. 61228202, and No. 61272426, China 973 Program under Grant No.2011CB302705, China Postdoctoral Science Foundation funded project under grant No. 2012M510029, NSF CNS-0832120, NSF CNS-1035894, NSF ECCS-1247944.}

\vspace{-10pt}
\section{Introduction}\label{section:introduction}

Great amount of data is collected for the information-based decision
making in our real life these days (Smart Grid
\cite{danezis2013smart}, Social Network Services
\cite{clemons2007future}, Location Based Services
\cite{xie2014location} \textit{etc.}). This trend originated from the
business needs. Enterprises have been increasingly looking to discover the
answers to specific business questions from the data they possess, and
various data mining techniques are applied to derive the answers
(\textit{e.g.,} marketing \cite{berry1997data} and health care
\cite{jadad2000internet}). Driven by such needs, publication, sharing
and analysis of user-generated data has become common. For
examples, President Information Technology Advisory Committee (PITAC)
released a report to establish a nation-wide electronic medical
records sharing system to share the medical knowledge for various
clinic decision making \cite{president2004revolutionizing}; AOL
published its anonymized query logs for research purpose
\cite{barbaro2006face}; Netflix released anonymized dataset containing
500,000 subscribers' movie ratings to organize a public contest for
the best recommendation algorithm \cite{bennett2007netflix}. However,
user-generated data in its original form contains much sensitive
information about individuals
(\cite{akcora2012privacy,danezis2009wisdom}), and the publication of
such data has brought active discussions in both industry and academia
\cite{carminati2013share,chow2009two,li2013scalable}. Current primary
practice is to rely on 1) sanitization to clean up the sensitive
information in the publication \cite{fung2010privacy}, or 2)
regulation enforcement to restrict the usage of the published data
\cite{gillin2000federal,hans2012privacy}. The former approach may
bring distortion to the data during the sanitization which leads to
accuracy loss, and the latter one solely relies on the legal or
voluntary agreements that the data should be used only for agreed purposes. The drawback of those protections is two folds. 1) A good
trade-off between the utility and the privacy is hard to find; 2) Even
if data recipients do not misbehave on the data, such protection does
not guarantee the protection, in the sense that a profitable sensitive
data may become the target of various types of attacks, and the
contracts or agreements cannot guarantee any protection in this case. 

Various previous works focus on the data anonymization
\cite{li2010closeness,mohaisen2013dynamix} to protect the identities of
the data owners, but the de-anonymization
\cite{danezis2013you,farenzena2010person} is its
powerful countermeasure, and it cannot provably guarantee individual
privacy. In this paper, we focus on a more substantial aspect of 
privacy implications in the third-party data analysis -- the
confidentiality of the data itself. We investigate the privacy
preserving data analysis problem in which a third-party aggregator
wants to perform data analysis on a private dataset generated by any subgroup of total users in the system, where the total users in the system may change dynamically. To
enable this without privacy concerns, it is desirable to have a data
analysis tool which can provably guarantee the data privacy without bringing distortion to the analysis result such that the aggregator
will get the most utility out of it. Various types of solutions based
on different techniques are proposed to support such privacy
preserving computation (secure multi-party computation, functional
encryption, perturbation \textit{etc.}, all reviewed in Section
\ref{section:related}), but we have the following design goals to make
our scheme applicable in a realistic environment, and most of existing
solutions fail to achieve one or more
goals.\vspace{-10pt} 

\subsection*{Design Goals}
\textbf{Varieties}: Two varieties need to be considered. In reality, the actual group of users generating data changes rapidly along the time, and the type of analysis itself also changes depending on the characteristics of the dataset. Therefore, a scheme needs to adapt itself to such two varieties.\\
\textbf{Time-series Data}: Users continuously generate data, therefore the data analysis scheme should preserve the data confidentiality even when the data is a time series and may be substantially self-correlated.\\
\textbf{Communication \& Computation Overhead}: Considering the rate at which data is generated and collected these days, the analysis on the published data should be efficient in terms of both computation and communication overhead.\\
\textbf{Channel Security}: In dynamic environments such as VANET, establishing pair-wise secure channels is almost impossible due to the huge overhead. Therefore a scheme cannot rely on secure channels to achieve privacy.\\
\textbf{Privacy-requirement Independent Accuracy}: Distorted result is not accepted in many cases, therefore we require accurate result with zero or negligible distortion from the published data, which is not a function of the guaranteed privacy level. This property will enable us to guarantee the proposed security (IND-CPA) all the time, hence trade-off between the security and utility is unnecessary.

In this paper, we design a framework for Privacy-preserving Data analytics (\ourprotocol), a cryptographic tool which enables third-party analysis on private data records without privacy leakages. The main contributions of this paper are summarized as follows.
\begin{compactenum}
\item Instead of presenting ad-hoc solutions for many applications, \ourprotocol~enables any type of data analysis that is based on polynomials (\textit{e.g.}, statistics calculation \cite{statistics_formula}, regression analysis, and support vector machine classification) to be performed on a private dataset distributed among individuals.

\item We propose a provably secure cryptographic framework which guarantees semantic security against chosen-plaintext attack. Specifically, any polynomial adversary's advantage is polynomially bounded by the advantage of solving a Decional Deffie-Hellman problem.

\item Although we support many different formats of polynomial-based data analysis among any subset of $n$ users, every user needs to hold only $O(n)$ keys, and the computation \& communication complexities for adding an extra user to existing group of $n$ users are both $O(n)$ while removing an existing user does not require extra operation. Besides, on average the communication overhead for an ordinary user is less than 1 KB per product term in the polynomial.
\end{compactenum}

%Furthermore, our methods are proposed for computing the value of a multi-variate polynomial function where the input of each participant is assumed to be an integer, our methods can be trivially generalized for functions (such as inner product) where the input of each participant is a vector.

%The rest of the paper is organized as follows. We compare our approach with existing works in Section \ref{section:related}, and formally define the problem and the protocol in Section \ref{section:problem}. The arithmetic protocol is presented in Section \ref{section:design}, and it is used to implement data aggregation scheme in Section \ref{section:scheme}. We analyze our protocol's security and performance in Section \ref{section:security} and Section \ref{section:performance}, and finally conclude in Section \ref{section:conclusion}. 

\section{Preliminaries}\label{section:problem}
\subsection{Notations}\label{sec:notations}

We use $\mathbb{Z}_N$ to denote the additive group $\{0,1,\cdots,N-1\}$, and $\mathbb{Z}_N^*$ to denote the invertible residues modulo $N$. Also, $|\mathbb{G}|$ and $|\mathcal{P}|$ are used to denote the order of group $\mathbb{G}$ and the size of the set $\mathcal{P}$ respectively. Recall that, given the conventional notation of Euler's totient function $\phi(N)$, $|\mathbb{Z}_N^*|=\phi(N)$.
When demonstrating the reducibility between the computation problems $X,Y$, we use $X\leq_P Y$ to denote that $X$ is polynomially reducible to $Y$ , and $X\equiv Y$ to denote that $X$ is polynomially equivalent to $Y$.

%\subsubsection{Invertible Residues Modulo $N$}
%
%We denote $\mathbb{Z}_N^*$ the multiplicative group of invertible residues of $\mathbb{Z}_N^*$ (\textit{i.e.}, $\{x\in\mathbb{Z}_N|\exists y\in\mathbb{Z}_N: xy=1\mod N\}$). Recall that $x^{\phi(N)}=1\mod N$ for any $x\in\mathbb{Z}_{N}^*$ due to the Euler's Theorem.\vspace{-10pt}
%\begin{definition}
%Decisional Diffie-Hellman (DDH) problem in the group $\mathbb{Z}_N^*$ with generator $g$ is to decide whether $g^c=g^{ab}$ given a triple $(g^a,g^b,g^c)$. An algorithm $\mathcal{A}$'s advantage in solving the DDH problem is defined as
%\begin{displaymath}\begin{split}
%adv_\mathcal{A}^{DDH}&=\Big|\mathbf{Pr}\big[1\leftarrow\mathcal{A}(\mathbb{Z}_N^*,g^a,g^b,g^{ab}\in\mathbb{Z}_N^*)\big]\\
%&-\mathbf{Pr}\big[1\leftarrow\mathcal{A}(\mathbb{Z}_N^*, g^a,g^b,g^c\in\mathbb{G})\big|c\leftarrow_R \mathbb{Z} \big]\Big|
%\end{split}\end{displaymath}
%where $1\leftarrow \mathcal{A}(\cdot)$ if the algorithm outputs `yes' and 0 otherwise, and the probabilities are taken over the random selection $c\leftarrow_R \mathbb{Z}$ as well as the random bits of $\mathcal{A}$.
%\end{definition}
%The group $\mathbb{Z}_N^*$ as an input to the algorithm $\mathcal{A}$ means the public description of the group (\textit{e.g.} generator $g$, moduli $N$) is given as an input.
\begin{definition}
$k$-th Decisional Diffie-Hellman ($k$-DDH) problem in the group $\mathbb{Z}_N^*$ with generator $g$ is to decide whether $(g^k)^c=(g^k)^{ab}$ given a triple $((g^k)^a,(g^k)^b,(g^k)^c)$, and any algorithm $\mathcal{A}$'s advantage in solving the $k$-DDH problem is denoted by $adv_{\mathcal{A}}^{k-DDH}$.
\end{definition}

%
%\subsubsection{$N$-th Residues Modulo $N^2$}\label{sec:N-residue}
%
%We use $\mathbb{NR}_{N^2}$ to represent the subgroup of $N$-th residues of $\mathbb{Z}_{N^2}^*$ (\textit{i.e.}, $\{x\in\mathbb{Z}_{N^2}^*|\exists y\in\mathbb{Z}_{N^2}^*:x=y^N\mod N^2\}$).
\begin{definition}
A number $w$ is called $N$-th residue modulo $N^2$ if there exists $y\in\mathbb{Z}_{N^2}^*$ such that $w=y^N=\mod N^2$.
\end{definition}
Recall that the set of $N$-th residues modulo $N^2$ is a cyclic multiplicative subgroup of $\mathbb{Z}_{N^2}^*$ of order $\phi(N)$. Every $N$-th residue has $N$ roots, and specifically, the set of all $N$-th roots of the identity element forms a cyclic multiplicative group $\{(1+N)^x=1+xN\mod N^2|x\in \mathbb{Z}_N\}$ of order $N$.
\begin{lemma}\label{lem:DLsolver}
There exists a function
\begin{displaymath}
{D}:\{(1+N)^x\mod N^2|x\in\mathbb{Z}_N\}\rightarrow \mathbb{Z}_N
\end{displaymath}
which can efficiently solve the discrete logarithm within the subgroup of $\mathbb{Z}_{N^2}^*$ generated by $g=1+N$.
\end{lemma}
\begin{proof}
For any $y=(1+N)^x$, $y=(1+xN)\mod N^2$. Then, the function $\mathcal{D}(y)=(y-1)/N=x$ efficiently solves the discrete logarithm within the subgroup, where `$(y-1)/N$' is the quotient of the integer division $(y-1)/N$ rather than $y-1$ times the inverse of $N$.
\end{proof}
\begin{definition}
Decisional Composite Residuosity (DCR) problem in the group of $N$-th residues modulo $N^2$ with generator $g$ is to decide whether a given element $x\in\mathbb{NR}_{N^2}$ is an $N$-th residue modulo $N^2$.
\end{definition}

\subsection{Computational Assumptions}\label{sec:comp_assumption}

$N$ is an RSA modulus of two safe prime numbers $p,q$ ($N=pq$, and $p,q$ satisfy $p=2p'+1,q=2q'+1$ for some large prime numbers $p',q'$ \cite{camenisch1999proving}). Then, the DDH assumption states that if the factorization of $N$ is unknown, the DDH problem in $\mathbb{Z}_{N}^*$ is intractable \cite{boneh1998decision,goldwasser1996lecture}. That is, for any probabilistic polynomial time algorithm (PPTA)  $\mathcal{A}$, $adv_\mathcal{A}^{DDH}$ is a negligible function of the security parameter $\kappa$.
\begin{lemma}\label{lem:kddh}
When $N$ is a semiprime of two safe prime numbers $p,q$ and the factorization $N=pq$ remains unknown,  we have $\text{DDH}\leq_P\text{$k$-DDH}$.
\end{lemma}
\begin{proof}
Given an oracle solves $k$-DDH problem, if one knows $k$, he can submit $((g^a)^k, (g^b)^k, (g^c)^{k})$ to this oracle. $g^c=g^{ab}$ if and only if the oracle outputs 1.
\end{proof}

%The above lemma shows $k$-DDH is believed to be intractable as well. Note that $\text{$k$-DDH}\leq_P\text{DDH}$ is not true unless $k^{-1}\mod \phi(N)$ is known. Since the factorization $N=pq$ is assumed to be unknown, $k^{-1}\mod \phi(N)$ remains unknown as well, and it is not possible to solve the DDH problem with the oracle solving $k$-DDH problem.

Under the same condition, the DCR assumption states that the DCR problem in $\mathbb{Z}_{N^2}^*$ is believed to be intractable if the factorization of $N=pq$ is unknown \cite{paillier1999public}. That is, for any PPTA $\mathcal{A}$, $adv_\mathcal{A}^{DCR}$ is a negligible function of the security pararmeter. 
In this paper, we rely on the above assumptions to define and prove the security.

\section{Problem Definitions}

\subsection{Problem Modelling}

\textit{W.l.o.g.}, we assume there are total $n$ \textbf{users}, and each of them is represented by an integer ID:$i\in\{1,\cdots,n\}$. Each user generates time-series data records $x_{i,t}$ at time $t$, but we use $x_{i}$ to denote his data value for visual comfort. A third-party \textbf{aggregator} wishes to choose an arbitrary subset of users $\mathcal{P}\subseteq \{1,\cdots,n\}$ to conduct certain data analysis on their private data records by evaluating:
\begin{equation}\label{equation:polynomial}
f(\mathbf{x}_{\mathcal{P}})=\sum_{k} c_{k}\Big(\prod_{i\in\mathcal{P}} {x_{ik}^{e_{ik}}}\Big)
\end{equation}
where $x_{ik}$ is user $i$'s private value for the $k$-th product term $\prod x_{ik}^{e_{ik}}$, and we call $\mathcal{P}$ as the participants of the polynomial evaluation. Note that Eq.(\ref{equation:polynomial}) can be used to represent any general multivariate polynomial because the coefficients $c_k$'s and the exponents $e_{ik}$'s are tunable parameters.

To model the time-variation, we use a discrete time domain $T=\{t_1,t_2,\cdots\}$ of time slots, and when the aggregator submits the query polynomial to users, he will also declare a sequential time window $T_f\subset T$ associated with the polynomial in order to request users' data generated during the time window $T_f$.

\subsection{Informal Threat Assumptions}

Due to privacy concerns, the analytic result should be only given to the aggregator, and any user's data should be kept secret to anyone else but the owner unless it is deducible from the polynomial value. Besides, both users and the aggregator are assumed to be  \textbf{passive} and \textbf{adaptive} adversaries. They are passive in that they will not tamper the correct computation of the protocol. If users tamper the protocol, it is highly likely that the aggregator will detect it since the outcome of the protocol will not be in a valid range (\textit{e.g.,} mean age$=3,671.9$) due to the cryptographic operations on large integers, and the aggregator is interested in the correct analytic result, so they will not tamper the protocol itself. Furthermore, users may report a value with small deviation such that the analytic result still `looks' good due to many reasons (moral hazards, unintentional mistake,  \textit{etc.}), but evaluating the reliability of the reported value is out of this paper's scope. On the other hand, they are adaptive in that they may produce their public values adaptively after seeing others' public values (\textit{i.e.,} passive rushing attack \cite{feigenbaum1991distributed}). Besides, we also assume the users will not collude with each other or the aggregator in our basic construction, and this assumption will be relaxed in the advanced version in the later section.

As aforementioned, we assume the Dolev-Yao network model. That is, all the communication channels are open to anyone, and anyone can overhear and synthesize any message. This is a reasonable assumption since in a dynamic network environment in which actual group of available users may change (\textit{e.g.,} sensor networks, crowdsourcing environment), establishing pairwise secure communication channels is costly and may not be possible due to limited resources.

\subsection{Privacy-preserving Data Analytics: \ourprotocol}\label{sec:definitions}

To formally define the correctness and the security of our framework, we first present a precise definition of \ourprotocol.

\begin{definition}
Our Privacy-preserving Data Analytics\\ (\ourprotocol) is the collection of the following four polynomial time algorithms: $\mathsf{Setup}$, $\mathsf{KeyGen}$, $\mathsf{Encode}$, and $\mathsf{Aggregate}$.\\
$\mathsf{Setup}(1^\kappa)\rightarrow \mathsf{params}$ is a probabilistic setup algorithm that is run by a crypto server to generate system-wide public parameters $\mathsf{params}$ given a security parameter $\kappa$ as input.\\
$\mathsf{KeyGen}(\mathsf{params})\xrightarrow{dist.}\{\SK_1,\SK_2,\cdots,\SK_n\}$ is a probabilistic and distributed algorithm jointly run by the users. Each user $i$ will secretly receive his own secret key $\SK_i$.\\
$\mathsf{Encode}(x_{ik},f(\cdot),\SK_i,T_f)\rightarrow C(x_{ik})$ is a deterministic algorithm run by each user $i$ to encode his private value $x_{ik}$ into a communication string $C(x_{ik})$ using $t_k\in T_f$. The output $C(x_{ik})$ is published in an insecure channel.\\
$\mathsf{Aggregate}(\{C(x_{ik})|\forall k,\forall i\in\mathcal{P}\})\rightarrow f(\mathbf{x}_\mathcal{P})=\sum(c_k\prod x_{ik}^{e_{ik}})$ is run by the aggregator to aggregates all the encoded private values $\{C(x_{ik})\}$'s to calculate $f(\mathbf{x}_\mathcal{P})$.
\end{definition}
We use `\textsf{Encode}' and `\textsf{Aggregate}' instead of conventional terminologies `\textsf{Encrypt}' and `\textsf{Decrypt}' because we will employ a homomorphic encryption as a black-box algorithm later.

\begin{definition}
\ourprotocol~is correct if for all $\mathcal{P}\subseteq\{1,\cdots,n\}$:
\begin{displaymath}\footnotesize
\mathbf{Pr}\left[
\begin{array}{c}
\mathsf{params}\leftarrow\mathsf{Setup}(1^\kappa);\\
\{EK_i\}_{i=1}^n\leftarrow\mathsf{KeyGen}(\mathsf{params});\\
\forall k,\forall i:C(x_{ik})\leftarrow\mathsf{Encode}(x_{ik},f(\mathbf{x}_\mathcal{P}),\SK_i,T_f);\\
\mathsf{Aggregate}(\{C(x_{ik})|\forall k,\forall i\})=f(\mathbf{x}_\mathcal{P})
\end{array}
\right]=1
\end{displaymath}
where the probabilities are taken over all random bits consumed by the probabilistic algorithms.
\end{definition}

The security of \ourprotocol~is formally defined via following data publishing game.\smallskip

\noindent\textbf{Data Analysis Game}:

\textsf{Setup}: 3 disjoint time domains are chosen: $T_1$ for phase 1, $T_2$ for phase 2, and $T_c$ for the challenge phase.

\textsf{Init}: The adversary declares his role in the \ourprotocol~scheme (\textit{i.e.,} aggregator or user), and  the challenger controls the remaining participants. Both of them engage themselves in the $\mathsf{Setup}$ and $\mathsf{KeyGen}$ algorithms.

\textsf{Phase 1 in $T_1$}: The adversary submits polynomially many calls to the encode oracle\footnote{The oracle generates and gives the communication strings at the adversary's request based on the queried polynomial $f(\mathbf{x}_\mathcal{P})$, its associated time window $T_f$, and the input values $\{x_{ik}\}_{i\in\mathcal{P},k}$. Challenger can act as such encode oracle in this game for the users that are not under adversary's control.} based on arbitrary query $(f(\mathbf{x}),T_f\subseteq T_1, \{x_{ik}\}_{i\in\mathcal{P},k})$. If the declared time windows do not overlap with each other, he receives all corresponding communication strings from the oracle. Otherwise, he receives nothing.

\textsf{Challenge in $T_c$}: The adversary declares the target polynomial $f^*(\mathbf{x}_\mathcal{P})$ to be challenged as well as its time window $T_{f^*}\subseteq T_c$. Then, he submits two sets of values $\mathbf{x}_{\mathcal{P}1},\mathbf{x}_{\mathcal{P},2}$, such that $f^*(\mathbf{x}_{\mathcal{P},1})=
f^*(\mathbf{x}_{\mathcal{P},2})$, to the challenger. The challenger flips a fair binary coin $b$ and generate the corresponding communication strings based on $\mathbf{x}_{\mathcal{P},b}$, which are given to the adversary.

\textsf{Phase 2 in $T_2$}: Phase 1 is repeated adaptively, but the time window $T_f$ should be chosen from $T_2$.

\textsf{Guess}: The adversary gives a guess $b'$ on $b$.\smallskip

The advantage of an algorithm $\mathcal{A}$ in this game is defined as $adv_\mathcal{A}^{\ourprotocol}=\big|\mathbf{Pr}[b'=b]-\frac{1}{2}\big|$. Intuitively, we assumed the worst-case scenario (\textit{i.e.,} most powerful attacker) in the above game where even an adversarial user can also specify the target polynomial to be evaluated, and the polynomial can be chosen adaptively to increase the advantage. In real-life applications, polynomials are chosen by the aggregator only, and the choice will be monitored by the users since the coefficients and powers are public.

\begin{definition}
Our \ourprotocol~is indistinguishable against chosen-plaintext attack (IND-CPA) if all polynomial time adversaries' advantages in the above game are of a negligible function \textit{w.r.t.} the security parameter $\kappa$ when $T_1,T_2$, and $T_c$ are three disjoint time domains.
\end{definition}

Note that our security definition does not guarantee that the private values cannot be deduced from the output values. Even a perfectly secure black-box algorithm cannot guarantee it if the same set of $k$ values are fed as inputs for $k$ different polynomials' evaluation (a simple linear system). Rather, we use IND-CPA to define the security which is shown to be equivalent to the semantic security \cite{goldwasser1984probabilistic}. In other words, our framework guarantees that any adversary's knowledge (either adversarial aggregator or user) on the probability distribution of the private values remains unchanged even if he has access to all the communication strings generated by our \ourprotocol. The indistinguishability becomes particularly crucial when the data is generated continuously along the time, and the same user's data may be temporally correlated. Unless guaranteed by the semantic security, the data privacy is hardly preservable.

%With the formal definitions above, our problem is to design the framework \ourprotocol~which is both correct and indistinguishable against CPA.

\section{Achieving Correct \& Secure \ourprotocol}

In this section, we first present a basic construction without colluding users for simplicity, and further presents the countermeasure to collusion attacks in Section \ref{section:attacks}.

\subsection{Intuitions}
The following two steps show the big sketch of our design on the oblivious polynomial evaluation.\\
\textbf{Hiding each value}: To hide every private value $x_{ik}\in\mathbb{Z}_N$ in the polynomial $f(\mathbf{x}_\mathcal{P})$ (Eq.\ref{equation:polynomial}), we let each user $i$ publish the perturbed data $x_{ik}r_{ik}$, where $r_{ik}$ is an independently chosen random noise, but it satisfies that $\prod_{i\in\mathcal{P}} r_{ik}=1$ with modulo operation using our novel secret sharing technique. It follows that the product of all perturbed data is equal to $\prod_i x_{ik}$ for all $k$.\\
\textbf{Hiding each product term}:  
To hide the product terms in Eq. (\ref{equation:polynomial}), we employ two users with special roles. The first special user $i_1$ will publish homomorphically encrypted perturbed data $E(x_{i_1k}r_{i_1k})$ instead of his perturbed data using the aggregator's key. Since the plaintext of this ciphertext is a perturbed data, the aggregator does not learn $x_{i_1k}$. On the other hand, because only the aggregator has the private key of the ciphertext, no one else learns $x_{i_1k}$ or $\prod_i x_{ik}$ either\footnote{Not all $x_ir_i$ terms are published, therefore the product of published perturbed data does not yield $\prod_i x_{ik}$}. The second user $i_2$ does not publish his perturbed data either. Instead, he conducts homomorphic operations:\vspace{-3pt}
\begin{displaymath}
E(x_{i_1k}r_{i_1k})^{\prod_{i\neq i_1}x_{ik}r_{ik}}=E(\prod_{i}x_{ik})\vspace{-4pt}
\end{displaymath}
for all $k$, where $E(m)$ denotes the ciphertext of $m$ encrypted with a homomorphic encryption scheme. Then, he locally picks random numbers $\mathcal{K}_k$ such that $\prod_k \mathcal{K}_k=1$ (modulo operation), and publishes perturbed ciphertexts $\mathcal{K}_kE(\prod_{i}x_{ik})$ to the aggregator. The product of them is equal to $E(\sum_k\prod_ix_{ik})$, and only the aggregator can decrypt it because other users do not have the private key. On the other hand, the aggregator cannot obtain values of any single product term because the ciphertexts are perturbed.

Letting the users independently obtaining random noises $r_{ik}$ such that $\prod_{i\in\mathcal{P}} r_{ik}=1$ is the most challenging part. To do so, we exploit the sequential time window $T_f$ associated with the polynomial $f(\cdot)$. All users will pick the $k$-th time slot $t_k$ in $T_f$ to generate $r_{ik}$ for his $x_{ik}$. Since the random noise is based on $t_k$, we need different time slots for all product terms. This implies we have to consume $T_f$ of size $m$ for a polynomial $f$ with $m$ product terms. Also, all sequential time windows for the evaluation must be disjoint to each other to guarantee that no time slots are used for twice.\vspace{-5pt}

\subsection{Constructions}\label{section:design}

Firstly, we define the Lagrange coefficient for $i\in\mathbb{Z}_n$ and a set, $\mathcal{P}$, of integers in $\mathbb{Z}_n$ as $
\mathcal{L}_{i,\mathcal{P}}(x)=\prod_{\substack{j\in \mathcal{P}\\j\neq i}}\frac{x-j}{i-j}$, and use the simplified Lagrange coefficient $\mathcal{L}_{i,\mathcal{P}}$ for a special case $\mathcal{L}_{i,\mathcal{P}}(0)$ for the visual comfort.

Then, our provably secure framework \ourprotocol~works as follows. Note that modular operations are omitted unless necessary for the sake of simplicity.\vspace{-5pt}

\subsubsection*{Setup}

A crypto server picks two safe semiprimes $N=pq$, $\tilde{N}=\tilde{p}\tilde{q}$ such that $p,q$ are of same length, $\tilde{p},\tilde{q}$ are of same length $\kappa$ bits, and $\phi(N)/\tilde{N}=k$ is an integer. Then, he randomly picks a generator $g$ of the group $\mathbb{Z}_N^*$, a generator $\tilde{g}$ of $\mathbb{Z}_{\tilde{N}}^*$, and sets $h=g^k$. Subsequently, he randomly chooses a hash function $H:\mathbb{Z}\rightarrow \langle h \rangle$, which is modelled as a random oracle. Finally, he publishes $\mathsf{params}=(g,\tilde{g},N,\tilde{N},H)$ and destroys $p,q,\tilde{p},\tilde{q}$ such that they are no longer available.
\begin{lemma}\label{lem:hN}
$h^{\tilde{N}}=1\mod N$.
\end{lemma}
\begin{proof}
$h^{\tilde{N}}=g^{k\tilde{N}}=g^{\phi(N)}=1\mod N$
\end{proof}\vspace{-3pt}

Any party independent from aggregator or users is qualify as the crypto server, for example, Internet Service Providers or governments.

\subsubsection*{Key Generation}

\textbf{Aggregator's Key Generation}: The aggregator generates a pair of additive homomorphic encryption keys $\HPK,\HSK$, and publishes $\HPK$ only. This additive homomorphic encryption must have the following properties:
\begin{displaymath}
\begin{array}{c}
E_\HPK(m_1)\cdot E_\HPK(m_2)=E_\HPK(m_1+m_2)\\
E_\HPK(m_1)^{m_2}=E_\HPK(m_1m_2)
\end{array}
\end{displaymath}
where $E_\HPK(m)$ denotes the ciphertext of $m$ encrypted with $\HPK$. Any additive homomorphic encryption with message space $\mathbb{Z}_N$ is accepted (\textit{e.g.,} Paillier's cryptosystem \cite{paillier1999public}).

\noindent \textbf{Users' Key Generation}: All users participate in the following secret sharing.
\begin{compactenum}
\item Every user $i$ ($i\in\{1,\cdots,n\}$) randomly picks $r_i\in\mathbb{Z}_{\tilde{N}}^*$, and broadcasts $y_i=\tilde{g}^{r_i}\in\mathbb{Z}_{\tilde{N}}^*$ via insecure channel.

\item Then, every user $i$ uses $y_{i+1}$ and $y_{i-1}$ to locally calculate $Y_i=(y_{i+1}\cdot y_{i-1}^{-1})^{r_i}\in\mathbb{Z}_{\tilde{N}}^*$, where the inverse is over modulo $\tilde{N}$. Specially, user 1 uses $y_2$ and $y_n$, and user $n$ uses $y_{1}$ and $y_{n-1}$.
\end{compactenum}
\begin{lemma}\label{lem:yN}
$\prod_{i=1}^n Y_i = 1\mod \tilde{N},\prod_{i=1}^n Y_i^{\tilde{N}}=1\mod {\tilde{N}}^2$
\end{lemma}
\begin{proof}\vspace{-3pt}
\begin{displaymath}\begin{split}\small
\prod_{i=1}^n Y_i=\tilde{g}^{(r_2-r_n)r_1}\cdots \tilde{g}^{(r_1-r_{n-1})r_n}
%&=\tilde{g}^{r_2r_1+r_3r_2+\cdots r_1r_n-r_nr_1-r_1r_2-\cdots r_{n-1}r_n}\mod \tilde{N}\\
=\tilde{g}^0=1\mod \tilde{N}
\end{split}\end{displaymath}
Then, $\prod_{i=1}^n Y_i=1+m\tilde{N}$ for some integer $m$, and $\prod_{i=1}^n Y_i^{\tilde{N}}=(1+m\tilde{N})^{\tilde{N}}=1\mod {\tilde{N}}^2$ as in Section \ref{sec:notations}.
\end{proof}

Subsequently, every user $j$ randomly defines a polynomial $q_j^{(d)}(x)$ over $\mathbb{Z}_{\tilde{N}}^*$, whose degree is $d$ and the constant term is 0 (\textit{i.e.,} $q_{j}^{(d)}(0)=0$). That is, every user randomly picks the coefficients for its own polynomial from $\mathbb{Z}_{\tilde{N}}^*$. This is repeated for all $d=2,\cdots,n-1$. Then, each user $i$ executes the encoding key query (Sub-Algorithm \ref{alg:key-query}), in which all other users jointly interact with him to respond to the query. The query outputs the specific data points on the secret global polynomials $q^{(2)}(x),\cdots, q^{(n-1)}(x)$ for each user, where the global polynomial is the sum of all individual polynomials.

\floatname{algorithm}{Sub-Algorithm}
\begin{algorithm}[h]
\caption{$\SK_i$ Query}\label{alg:key-query}
\begin{algorithmic}[1]
\STATE Every user $j\neq i$ computes and publishes
\begin{displaymath}
Q_{ij}=Y_j^{\tilde{N}}(1+\tilde{N})^{q_j^{(d)}(i)}\mod \tilde{N}^2
\end{displaymath}

\STATE After seeing all the broadcasts, user $i$ calculates:
\begin{displaymath}\small
\begin{split}
Y_i^{\tilde{N}}(1+\tilde{N})^{q_i^{(d)}(i)}\prod_{j\neq i}Q_{ij}=(1+\tilde{N})^{\sum_{j=1}^n q_j^{(d)}(i)}\mod \tilde{N}^2
\end{split}
\end{displaymath}
Subsequently, he uses aforementioned discrete logarithm solver $\mathcal{D}(\cdot)$ to solve $q^{(d)}(i)=\sum_j q_j^{(d)}(i)\mod \tilde{N}$ from it, which is set as one of his encoding keys. Later, $q^{(d)}(i)$ will be used to generate the data masker when $d+1$ users' data is to be analyzed.
\end{algorithmic}
Step 1 \& 2 are repeated for $d=2,\cdots,n-1$.\\
\textbf{Output}: $\SK_i(q^{(2)}(i),\cdots,q^{(n-1)}(i))$
\end{algorithm}
Essentially, in the user's key generation, each user $j$ randomly generates polynomials $q_j^{(2)}(x),\cdots,q_j^{(n-1)}(x)$ to respond others' key queries, and each key recipient $i$ receives one data point for each secret polynomial $q^{(d)}(i)$, where the polynomial $q^{(d)}(x)$ remains unknown to everyone because it is merely the sum of all users' random polynomials. However, specific point on this unknown polynomial can be jointly computed as shown in the query.
\begin{lemma}\label{lem:sum0}
For any set of users $\mathcal{P}$, we have:$$\sum_{i\in\mathcal{P}}q^{(|\mathcal{P}|-1)}(i)\mathcal{L}_{i,\mathcal{P}}=0\mod \tilde{N}$$
\end{lemma}
\begin{proof}
Recall that the constant terms in all polynomials are 0. Then, due to the polynomial interpolation, the above sum equals $q^{(|\mathcal{P}|-1)}(0)=0\mod \tilde{N}$.
\end{proof}\vspace{-3pt}

\subsubsection*{Input Encoding}

Recall that we employ two users with special roles. For the simplicity of the presentation, we assume that the first two users (user 1 \& 2) have to participate in the polynomial evaluation and act as the special users, but any other ordinary user can replace them.

After the aggregator declares the target polynomial $f(\mathbf{x}_{\mathcal{P}})=\sum_k c_k\big(\prod_i x_{ik}^{e_{ik}}\big)$ as well as the corresponding time window $T_f$, all users examine the time window. If $T_f$ overlaps with any of the consumed time windows, all users abort. Otherwise, every user $i\in\mathcal{P}$ except user 1 \& 2 computes and publishes his encoded value $C(x_{ik})$ for all $k$:
\begin{displaymath}
\forall k:C(x_{ik})=x_{ik}^{e_{ik}}H(t_k)^{q^{(|\mathcal{P}|-1)}(i)\mathcal{L}_{i,\mathcal{P}}}\in\mathbb{Z}_N^*\footnote{$C(x_{ik})$ may not be in $\mathbb{Z}_N^*$ when $x_{ik}$ is not invertible under modulo $N$, but this occurs only when $x_{ik}$ is not co-prime to $n$, \textit{i.e.,} $x_{ik}$ is equal to $p$ or $q$, two large safe prime numbers. The chance this happens is $\frac{1}{pq}$, and this is also the case where the large RSA number $n$ is factorized. We simply ignored this chance because it is a negligible function of the security parameter, and it is also commonly believed that the factorization is intractable.}
\end{displaymath}
where $t_k$ is the $k$-th time slot in $T_f$. At the same time, user 2 computes and publishes:\vspace{-2pt}
\begin{displaymath}
\forall k:E_{\HPK}\Big(C(x_{2k})\Big)\in\mathbb{Z}_{N^2}^*
\end{displaymath}
Subsequently, user 1 computes $C(x_{1k})$'s for himself and computes all $V_k$'s for the product terms as shown below:
\begin{displaymath}\begin{split}
\forall k:V_k&=E_{\HPK}\Big(C(x_{2k}^{e_{2k}})\Big)^{\prod_{i\in\mathcal{P},i\neq 2}C(x_{ik}^{e_{ik}})}=E_\HPK\Big(\prod_{i\in\mathcal{P}}x_{ik}^{e_{ik}}\Big)
\end{split}\end{displaymath}
Subsequently, he randomly selects $\MK_k\in\mathbb{Z}_{N^2}^*$ such that $\prod_k \MK_k=1\mod N^2$, and he publishes $\MK_kV_k^{c_k}$ for all $k$.

Even if a user $i$'s $x_{ik}$ does not appear in a product term $\prod x_{ik}^{e_{ik}}$, he still sets $x_{ik}=1$ and participates in the analysis by following the algorithm to guarantee the correctness of the final aggregation.\vspace{-3pt}

\subsubsection*{Aggregation}

When all users finish their encoding and broadcasting, the aggregator conducts the following homomorphic operations:\vspace{-6pt}
\begin{displaymath}\begin{split}
\prod_k \MK_kV_k^{c_k}&=E_\HPK\Bigg(\sum_kc_k\Big(\prod_{i\in\mathcal{P}}x_{ik}^{e_{ik}}\Big)\Bigg)\mod N^2
\end{split}\end{displaymath}
Then, he can use $\HSK$ to decrypt the polynomial value.

analysis in the appendix show that \ourprotocol~is correct and indistinguishable against chosen-plaintext attack. Note that the lowest degree in a user's encoding key $\SK_i$ is 2, and this forces at least three users to participate in the polynomial evaluation. This is because our scheme cannot guarantee the data privacy against other users when only two users participate in the evaluation (Section \ref{section:limit}).\vspace{-5pt}

\subsection{Supporting Dynamic User Group}
\label{section:dynamic_group}

Up to this point, our design is able to support data analysis in dynamic subgroup of a fixed group of $n$ users. A more substantial ability is to support a dynamic group of users, \textit{i.e.,} to enable new users to join the system and old users to leave the system.

\subsubsection*{Adding a new user $n+1$}

When a new user whose ID is $n+1$ needs to join the group of users $\{1,\cdots,n\}$, the number of participants in one round of data analysis can be as large as $n+1$. The first step to do is to update the key set of every user $i$ in the old group $\{1,\cdots,n\}$: each user $i$ needs to acquire and add one extra item $q^{(n)}(i)$ into his key set $\SK_i$ for the worst case where all of $n+1$ users participate in the analysis (so that their $q^{(n)}(i)$'s can be used when such case occurs). The second step is to issue the new user $n+1$ his full key set $\SK_{n+1}=\{q^{(2)}(n+1),\cdots,q^{(n-1)}(n+1),q^{(n)}(n+1)\}$ so that he can participate in any data analysis consisting of $2,3,\cdots,n+1$ participants.

For the first step, all users in the new group perform the aforementioned secret sharing in the Key Generation, and every user $j\in\{1,\cdots,n+1\}$  randomly defines $q_j^{(n)}(x)$ whose degree is $n$ and the constant term is 0. Then, for each user $i\in\{1,\cdots,n+1\}$, every user $j\neq i$ computes and publishes $Q_{ij}$ of degree $n$ as in Sub-algorithm \ref{alg:key-query}, and $i$ calculates $q^{(n)}(i)$ in the same way. When this is repeated for all users, every user $i$ in the old group $\{1,\cdots,n\}$ has a complete key set which contains key items $q^{(2)}(i),\cdots,q^{(n)}(i)$, and the new user $n+1$ only has $q^{(n)}(n+1)$.

For the second step, every user $j\neq i$ performs the following process for $d=2,\cdots,n-1$. User $j$ re-uses his previously defined individual polynomial $q_j^{(d)}(x)$ to calculate $q_j^{(d)}(n+1)$ and publishes  $Q_{n+1,j}$ of degree $d$ as in Sub-algorithm \ref{alg:key-query}. Subsequently, user $n+1$ calculates the product and acquire the $q^{(d)}(n+1)$ in the same way. When this is repeated for $d=2,\cdots,n-1$, the new user acquires $q^{(2)}(n+1),\cdots,q^{(n-1)}(n+1)$.

Finally, $2n+1$ new items are shared among $n+1$ users, and every user $i$ in the new group has his own $n+1$ items in the key set $\SK_i$, which is the full key set for the new group of users $\{1,\cdots,n+1\}$.

\subsubsection*{Adding $n'$ new users}

Batch arrival can be handled in a similar way. When $n'$ new users arrive at the same time, one round of secret sharing occurs in the new group of $n+n'$ users. As the first step, $n'$ extra key items $q^{(n)}(i),\cdots,q^{(n+n'-1)}(i)$ are delivered to every $i\in\{1,\cdots,n+n'\}$. Then, as the second step, every new user $i\in\{n+1,\cdots,n+n'\}$ acquires $q^{(2)}(i),\cdots,q^{(n-1)}(i)$ as aforementioned.

\subsubsection*{Removing one or more existing users}

When one or more users are removed or leave the group, the system can simply ignore the removed users, and  keys of the rest users are not affected by the removal.

\subsection{Representing Real Numbers}\label{section:integer_real}

Due to the cryptographic operations, all the computations and operations in our scheme are closed in integer groups, and thus we need to use integers to represent real numbers. We exploit the homomorphism in \ourprotocol~to represent real numbers via integers using the fixed point representation \cite{nikolaenko2013privacy}. Given a real number $x$, its fixed point representation is given by $[x]=\lfloor x\cdot 2^e \rfloor$ for a fixed $e$, and all the arithmetic operations reduce to the integer versions as follows:
\begin{compactitem}
\item Addition/Subtraction: $[x\pm y]=[x]\pm [y]$

\item Multiplication/Division : $[x\cdot y^{\pm 1}]=[x]\cdot [y]^{\pm 1}\cdot 2^{\mp e}$
\end{compactitem}
where $x\cdot 2^{-1}$ stands for $x$ divided by 2. Then, we have the following conversions for the arithmetic operations used in our schemes if using fixed point representations.
\begin{compactitem}
\item $C([x])C([y])$ $\Rightarrow$ $C([x])C([y])\cdot 2^{-e}$.

\item $E_\HPK([x])^{[y]}$ $\Rightarrow$ $E_\HPK([x])^{[y] \cdot 2^{-e}}$
\end{compactitem}
where $C(m)$ is the encoded communication string of $[m]$ (\textsf{Encode} algorithm), and $E_\HPK(m)$ denotes the ciphertext of $m$ encrypted with $\HPK$ (\textsf{Key Generation} algorithm). We do not use floating point representation because it is hard to implement the arithmetic operations on floating point numbers. Although this conversion gives practically high precision, one may consider employing \cite{aliasgari2013secure,franz2013secure} for much higher precisions.

\subsection{Data Analysis via \ourprotocol: Examples}\label{section:applications}

We only present some examples in this section since infinitely many different analysis are supported and it is not possible to enumerate all of them here, but any data analysis based on polynomials can be conducted via our framework.

\noindent \textbf{Statistics Calculation}: Various statistical values including, but not limited to mean, variance, sample skewness, mean square weighted deviation are all polynomials of the data values, and all of them can be privately evaluated by a third-party aggregator without knowing individual data by leveraging our \ourprotocol.

\noindent \textbf{Regression Analysis}:  Regression analysis is a statistical process for estimating the relationship among the variables. Widely used polynomial regression and ridge regression both belong to it. In both, each user $i$'s data record is described as a feature vector $\mathbf{x}$ and a dependent variable $y_i$, and training a regression model is to find $\mathbf{p}$ which minimizes $MSE(\mathbf{p})=\sum_i(y_i-\mathbf{p} \mathbf{x}_i)^2$, \textit{i.e.}, the linear predictor who predicts users' dependent variable vector $\mathbf{y}$ using their feature matrix $\mathbf{X}$ with minimum mean squared error. Since $MSE(\mathbf{p})$ is convex, it is minimized if and only if $\mathbf{A}\mathbf{p}=\mathbf{b}$, where $\mathbf{A}=\mathbf{X}^T\mathbf{X}$ and $\mathbf{b}=\mathbf{X}^T \mathbf{y}$. Further, $\mathbf{A}=\sum_i \mathbf{x}_i\mathbf{x}_i^T$, and $\mathbf{b}=\sum_i y_i\mathbf{x}_i$. Therefore, the aggregator can obliviously evaluate them with several calls to our \ourprotocol. Note that any polynomial regression model or ridge regression model can be trained in the same way.

\noindent \textbf{Support Vector Machine}: Support Vector Machine (SVM) is a supervised learning model which is widely used to classify the data records. Similar to the regression analysis, a user $i$'s data record is described as a feature vector $\mathbf{x}_i$ and its class $y_i\in\{-1,1\}$, and training a SVM is to find a vector $\mathbf{w}$ and a constant $b$ such that the hyperplane $\mathbf{w}\mathbf{x}-b=0$ separates the records having $y_i=-1$ from those having $y_i=1$ with the maximum margin $\frac{2}{|\mathbf{w}|}$. The dual form of this optimization problem is represented as maximizing $L(a)$ in $\mathbf{a}=\{a_1,\cdots,a_n\}$:\vspace{-3pt}
\begin{displaymath}\begin{split}
L(\mathbf{a})=\sum_{i=1}^na_i-\frac{1}{2}\sum_{i,j}&a_ia_jy_iy_j\mathbf{x}_i^T\mathbf{x}_j
\end{split}\end{displaymath}\vspace{-3pt}

\noindent subject to $\sum_{i=1}^na_iy_i=0$ and $0\leq a_i\leq C$. This is solved by the Sequential Minimum Optimization (SMO) algorithm \cite{platt1998sequential} with user-defined tolerance, who greedily seeks for the local maximum iteratively until convergence with certain error. Due to the equality constraint, $L(\mathbf{a})$ becomes a one-dimensional concave quadratic function in each iteration, which can be trivially optimized. The coefficients of the quadratic function are polynomials of all users' feature vectors and class values, and the aggregator can obliviously calculate them using our framework and optimize $L(\mathbf{a})$ at each iteration locally.

\section{Correctness \& Security Proofs}

\subsection{Correctness Proof}

\begin{theorem}\label{theorem:correct}
Our \ourprotocol~is correct.
\end{theorem}
\begin{proof}
We show by steps that each algorithm correctly calculates the output, which guarantees the correctness of our scheme.

\noindent \textit{Key Generation}: Correctness of the homomorphic encryption is already shown \cite{paillier1999public}, therefore we focus on the user's key generation. Due to Lemma \ref{lem:yN}, we have:
\begin{displaymath}
\begin{split}
Y_i^{\tilde{N}}(1+\tilde{N})^{q_i^{(d)}(i)}\prod_{j\neq i}Q_{ij}&=(\prod_{j=1}^n Y_j^{\tilde{N}})(1+\tilde{N})^{\sum_{j=1}^n q_j^{(d)}(i)}\\
&=(1+\tilde{N})^{\sum_{j=1}^n q_j^{(d)}(i)}\mod \tilde{N}^2
\end{split}
\end{displaymath}
Further, the existence of the discrete logarithm solver (Lemma \ref{lem:DLsolver}) guarantees that the discrete logarithm will be correctly calculated.

\textit{Input Encoding}\\
User 1 conducts the following computation in the input encoding:
\begin{displaymath}\begin{split}
\forall j:V_j&=E_{\HPK}\Big(C(x_2^{e_{2j}})\Big)^{\prod_{i\in\mathcal{P},i\neq 2}C(x_i^{e_{ij}})}\\
&=E_\HPK\Big(\prod_{i\in\mathcal{P}}C(x_i^{e_{ij}})\Big)\\
%&=E_\HPK\Big(H(t_k)^{\sum_{i\in\mathcal{P}}q^{(|\mathcal{P}|-1)}(i)\mathcal{L}_{i,\mathcal{P}}(0)}\prod_{i\in\mathcal{P}}x_{ik}^{e_{ik}}\Big)\\
&=E_\HPK\Big(H(t_k)^0\prod_{i\in\mathcal{P}}x_{ik}^{e_{ik}}\Big)=E_\HPK\Big(\prod_{i\in\mathcal{P}}x_{ik}^{e_{ik}}\Big)
\end{split}\end{displaymath}
The first two equations are true because of the additive homomorphism of the encryption $E_\HPK(\cdot)$, and the third equation is true because:
\begin{displaymath}\begin{split}
&\sum_{i\in\mathcal{P}}q^{(|\mathcal{P}|-1)}(i)\mathcal{L}_{i,\mathcal{P}}(0)=0\mod \tilde{N}~~~\text{(Lemma \ref{lem:sum0})}\\
&\Rightarrow H(t)^{\sum_{i\in\mathcal{P}}q^{(|\mathcal{P}|-1)}(i)\mathcal{L}_{i,\mathcal{P}}(0)}=(h^t)^0\mod N~\text{(Lemma \ref{lem:hN})}
\end{split}\end{displaymath}
%considering that the message space of the encryption $E_\HPK(\cdot)$ is $\mathbb{Z}_N$.
Correctness of the aggregation algorithm is obvious. In conclusion, after all $\mathsf{Setup,KeyGen,Encode}$ are executed, the output of $\mathsf{Aggregate}$ is $f(\mathbf{x}_\mathcal{P})$ with probability 1.
\end{proof}

\subsection{Security Proof}

To prove the indistinguishability (IND-CPA) of our \ourprotocol, we first prove the following lemmas.

\begin{lemma}
In the $\mathsf{KeyGen}$ algorithm, $Y_i$ is statistically indistinguishable from a uniformly random element chosen from $\mathbb{Z}_{\tilde{N}}^*$.
\end{lemma}
\begin{proof}
In the first secret-sharing phase, every user publishes $y_i=\tilde{g}^{r_i}$ where $r_i$ is randomly chosen, but since DDH problem is believed to be intractable in $\mathbb{Z}_{\tilde{N}}$ when the factorization of $\tilde{N}$ is unknown, $\tilde{g}^{r_i}$ is indistinguishable from a uniformly random element chosen from $\mathbb{Z}_{\tilde{N}}$ as shown in \cite{boneh1998decision}.
\end{proof}

\begin{lemma}
In the $\mathsf{KeyGen}$ algorithm, any secret data point $q_j^{(d)}(i)$ remains statistically indistinguishable from a uniform randomly chosen element, and all the encoding keys are indistinguishable from uniform randomly chosen elements.
\end{lemma}
\begin{proof}
In the $\SK_i$ Query (Sub-Algorithm \ref{alg:key-query}) phase, every user $i$ except user $j$ publishes $Q_{ij}=Y_j^{\tilde{N}}(1+\tilde{N})^{q_j^{(d)}(i)}$. We show that $Q_{ij}$ leaks no statistical information about the hidden random polynomial $q_j^{(d)}(i)$ as follows.

We first prove that, given $m$ and $y=Y_j^{\tilde{N}}(1+\tilde{N})^m$ only (except $Y_j$), deciding whether $m$ is the corresponding hidden value in $y$ (denoted as D-Hidden problem) is equivalent to the DCR problem. We suppose there is an oracle $\mathcal{O}_{D-Hidden}$ who can solve the D-Hidden problem, and another oracle $\mathcal{O}_{DCR}$ who can solve the DCR problem.

$\text{D-Hidden}\leq_P\text{DCR}$: Since $(1+\tilde{N})$ is an $\tilde{N}$-th non-residue, $y$ is an $\tilde{N}$-th residue if and only if $m=0\mod \tilde{N}$. Then, given $y$ and $m$, one can submit $y(1+\tilde{N})^{-m}$ to $\mathcal{O}_{DCR}$ to see if it is an $\tilde{N}$-th residue to decide whether $m$ is the hidden value.

$\text{DCR}\leq_P\text{D-Hidden}$: Similarly, given $x$ in the DCR problem (deciding if $x$ is an $\tilde{N}$-th residue), one can submit $m=0$ and $x$ to $\mathcal{O}_{D-Hidden}$ to see if $m=0$ is the corresponding hidden value in $x$ to see whether $x$ is an $\tilde{N}$-th residue or not.

Therefore, deciding whether $m$ is a hidden value in $y$ is exactly as hard as the DCR problem. Due to the same theory, given two known messages $m_0,m_1$ as well as $y=Y_j^{\tilde{N}}(1+\tilde{N})^{m_b}$, deciding which message is the corresponding hidden value is exactly as hard as the DCR problem, which concludes the semantic security of $Q_{ij}$ against chosen-plaintext attack.

Due to the semantic security of $Q_{ij}$, each user's individual polynomial value is statistically indistinguishable from a uniform random element, and therefore any encoding key $q^{(d)}(i)$ of any user $i$ is indistinguishable from an element uniform randomly chosen from $\mathbb{Z}_{\tilde{N}}^*$ since it's the sum of all individual polynomial values.
\end{proof}

With the above lemmas, we are ready to present the proof for the following theorem.
\begin{theorem}\label{theorem:security}
With the DDH assumption and DCR assumption, our \ourprotocol~scheme is indistinguishable against chosen-plaintext attack (IND-CPA) under the random oracle model. Namely, for any PPTA $\mathcal{A}$, its advantage $adv_\mathcal{A}^{\ourprotocol}$ in the Data Analysis Game is bounded as follows:
\begin{displaymath}
adv_\mathcal{A}^{\ourprotocol}\leq \frac{e(q_c+1)^2}{q_c}\cdot adv_{\mathcal{A}}^{DDH}
\end{displaymath}
where $e$ is the base of the natural logarithm and $q_{c}$ is the number of queries to the encode oracle.
\end{theorem}
\begin{proof}
To prove the theorem, we present three relevant games Game 1, Game 2, and Game 3, in which we use $\mathcal{A}$ and $\mathcal{B}$ to denote the adversary and the challenger. For each $l\in\{1,2,3\}$, we denote $E_l$ as the event that $\mathcal{B}$ outputs $1$ in the Game $l$, and we define $adv_l=|\mathbf{Pr}[E_l]-\frac{1}{2}|$.

\textbf{Game 1}: This is the game exactly identical to the Data Publishing Game in Section \ref{sec:definitions}. $\mathcal{A}$'s encode queries $(f(\mathbf{x})$, $T_f$, $\{x_{ik}\}_{i,k})$ are answered by returning the encode values $\{C(x_{ik})\}_{i,k}$. In the challenge phase, the adversary $\mathcal{A}$ selects a target polynomial $f^*(\mathbf{x})$, the associated time window $T_{f^*}$, and two set of values $\mathbf{x}_{\mathcal{P},1},\mathbf{x}_{\mathcal{P},2}$ which satisfy $f^*(\mathbf{x}_{\mathcal{P},1})= f^*(\mathbf{x}_{\mathcal{P},2})$. Then, the challenger $\mathcal{B}$ returns the corresponding encoded values to the adversary $\mathcal{A}$. When the game terminates, $\mathcal{B}$ outputs 1 if $b'=b$ and 0 otherwise. By the definition, $adv_1=|\mathbf{Pr}[E_1]-\frac{1}{2}|=adv_\mathcal{A}^{\ourprotocol}$.

\textbf{Game 2}: This game occurs after the Game 1 terminates. In Game 2, the adversary $\mathcal{A}$ and the challenger $\mathcal{B}$ repeat the same operations as in Game 1 using the same time windows of those operations. However, for each encode query in Game 1 at time $t\in T_1\cup T_2$, the challenger $\mathcal{B}$ flips a biased binary coin $\mu_{T_f}$ for the entire time window $T_f$ which takes 1 with probability $\frac{1}{q_c+1}$ and 0 with probability $\frac{q_c}{q_c+1}$. When the Game 2 terminates, $\mathcal{B}$ checks whether any $\mu_{T_f}=1$. If there is any, $\mathcal{B}$ outputs a random bit. Otherwise, $\mathcal{B}$ outputs 1 if $b'=b$ and 0 if $b'\neq b$. If we denote $F$ as the event that $\mu_{T_f}=1$ for any $T_f$, the same analysis as in \cite{coron2000exact} shows that $\mathbf{Pr}[\overline{F}]=\frac{1}{e(q_c+1)}$. According to \cite{joye2013scalable}, Game 1 to Game 2 is a transition based on a failure event of large probability, and therefore we have $adv_2=adv_1\mathbf{Pr}[\overline{F}]=\frac{adv_1}{e(q_c+1)}$.

\textbf{Game 3}: In this game, the adversary $\mathcal{A}$ and the challenger $\mathcal{B}$ repeat the same operations as in Game 1 using the same time windows of those operations. However, there is a change in the answers to the encode queries $(f(\mathbf{x}),T_f,\{x_{ik}\}_{i,k})$. The oracle will respond to the query with the following encoded values:
\begin{displaymath}
\forall i,\forall k:C(x_{ik})=\begin{cases}
x_{ik}^{e_{ik}}H(t_k)^{q^{(|\mathcal{P}|-1)}(i)\mathcal{L}_{i,\mathcal{P}}(0)} & \mu_{T_f}=1 \\
x_{ik}^{e_{ik}}\Big(H(t_k)^s\Big)^{q^{(|\mathcal{P}|-1)}(i)\mathcal{L}_{i,\mathcal{P}}(0)} & \mu_{T_f}=0
\end{cases}
\end{displaymath}
where $s$ is a uniform randomly chosen element from $\mathbb{Z}_{\tilde{N}}$ fixed for each polynomial $f(\mathbf{x})$, \textit{i.e.,} the same polynomial has the same $s$. When Game 3 terminates, $\mathcal{B}$ outputs 1 if $b'=b$ and 0 otherwise.

Due to the Lemma \ref{lem:distinguish_hard} below, distinguishing Game 3 from Game 2 is at least as hard as a DDH problem for any adversary $\mathcal{A}$ in Game 2 or Game 3. It follows then: $\Big|\mathbf{Pr}[E_2]-\mathbf{Pr}[E_3]\Big|\leq adv_\mathcal{A}^{DDH}$. The answers to encode queries in Game 3 are identical to those of Game 2 with probability $\frac{1}{q_c+1}$ and different  from those of Game 2 with probability $\frac{q_c}{q_c+1}$. In the latter case, due to the random element $s$, $C(x_{ik})$ is uniformly distributed in the subgroup of order $\tilde{N}$, which completely blinds $x_{ik}^{e_{ik}}$, and $\mathcal{A}$ can only randomly guess $b'$ unless he knows the factorization $N=pq$. Then, $b'=b$ with probability 1/2, and the total probability $\mathbf{Pr}[E_3]=\frac{\mathbf{Pr}[E_2]}{q_c+1}+\frac{q_c}{2(q_c+1)}$. Then, we have:
\begin{displaymath}
\begin{split}
\Big|\mathbf{Pr}[E_2]-\mathbf{Pr}[E_3]\Big|=&\Big|(\mathbf{Pr}[E_2]-1/2)\cdot\frac{q_c}{q_c+1} \Big|\\
=&adv_2\cdot \frac{q_c}{q_c+1}\leq adv_\mathcal{A}^{DDH}
\end{split}
\end{displaymath}
Combining the above inequality with the advantages deduced from Game 1 and Game 2, we finally have:
\begin{displaymath}
adv_2\cdot \frac{q_c}{q_c+1}=adv_\mathcal{A}^{\ourprotocol}\cdot \frac{q_c}{e(q_c+1)^2}\leq adv_\mathcal{A}^{DDH}
\end{displaymath}
\end{proof}

\begin{lemma}\label{lem:distinguish_hard}
Distinguishing Game 3 from Game 2 is at least as hard as solving a DDH problem for $\mathcal{A}$.
\end{lemma}
\begin{proof}
The only difference between Game 2 and Game 3 is the oracle's answers to the encode queries. When $\mu_{T_f}=1$, the answers in Game 3 are exactly same as the ones in Game 2. However, when the failure event $F$ occurs (\textit{i.e.}, $\mu_{T_f}=0$), $\mathcal{B}$ answers using $H(t)^{s}$ instead of $H(t)$. Since the values $x_{ik}$ is submitted as a query by $\mathcal{A}$, $\mathcal{A}$ knows the value of $x_{ik}$. Then, distinguishing Game 2 and Game 3 is equivalent to distinguishing the following two terms:
\begin{displaymath}
H(t)^{q^{(|\mathcal{P}|-1)}(i)\mathcal{L}_{i,\mathcal{P}}(0)}~~~\text{v.s.}~~~\Big(H(t)^s\Big)^{q^{(|\mathcal{P}|-1)}(i)\mathcal{L}_{i,\mathcal{P}}(0)}
\end{displaymath}
Since $H(t)$ maps $t\in\mathbb{Z}$ to a cyclic multiplicative group $\langle h\rangle$, we let $H(t)=h^{x}$ for some integer $x$. Let $a=x$ and $b=q^{(|\mathcal{P}|-1)}(i)\mathcal{L}_{i,\mathcal{P}}(0)$. Given a triple $(h^a, h^b, h^c)$, deciding whether $h^c$ is $h^{ab}$ or a random element in $\mathbb{Z}_N^*$ is exactly the $k$-DDH problem. Since $s$ is a uniform random number, distinguishing between the above two terms is equivalent to solving this $k$-DDH problem. However, in Game 3, the adversary is only given $h^a$ except $h^b$, which means distinguishing Game 3 from Game 2 is at least as hard as the $k$-DDH problem.

Due to Lemma \ref{lem:kddh}, the $k$-DDH problem is at least as hard as the DDH problem when $k$ is known. However, $k=\phi(N)/{\tilde{N}}$ is unknown to $\mathcal{A}$, and therefore the $k$-DDH in our group is also at least as hard as the DDH problem for $\mathcal{A}$.

In conclusion, distinguishing Game 3 from Game 2 is at least as hard a DDH problem for $\mathcal{A}$.
\end{proof}

\section{Performance Evaluation}\label{sec:evaluation}

\subsection{Theoretic Performance: Storage \& Communication}\label{sec:theoretic_analysis}

\noindent \textbf{Key Storage Complexity}: We only require every user to keep a set of encoding keys $\SK_i=\{q^{(2)}(i),\cdots,q^{(n-1)}(i)\}$. Lagrange coefficients $\mathcal{L}_{i,\mathcal{P}}$ needed in \textsf{Encode} can be locally computed when $\mathcal{P}$ of the analysis is declared. Therefore, the storage complexity is $O(n)$, where $n$ is the total number of users. To the best of our knowledge, there is no sub-linear solution under the same threat model and performance requirement (Section \ref{section:related}).

\noindent \textbf{Communication Overhead}: The communication overhead for evaluating a certain polynomial $f(\mathbf{x}_\mathcal{P})$ in terms of transmitted bits is summarized in the Table \ref{table:communication}. $\kappa$ is the bit-length of the safe prime numbers used in our scheme, $n'$ is the number of new users added into the system, $m$ is the total number of product terms in the polynomial (Eq. (\ref{equation:polynomial})), and $\theta_{min}$ is the minimum threshold of the number of participants in the evaluation.
\begin{table}[h]\small
\centering
\caption{Communication Overhead}\label{table:communication}\vspace{-10pt}
\begin{tabular}{r|c|c}
\hline\hline
Algorithm & Send (bits) & Receive (bits) \\\hline
 \multicolumn{3}{c}{Authority}\\\hline
\textsf{Aggregate} & 0 & $O(m\kappa)$ \\\hline
\multicolumn{3}{c}{Special User 1}\\\hline
\textsf{KeyGen when $k$ users collude} & $O(kn^2\kappa)$ &
 $O(kn^2\kappa)$\\
\textsf{Encode} & $O(m\kappa)$ & $O(\theta_{min}m\kappa)$ \\
\textsf{Adding $n'$ users} & $O(n'n)$ & $O(n'n)$ \\
\hline
\multicolumn{3}{c}{Other participants including Special User 2}\\\hline
\textsf{KeyGen when $k$ users collude} & $O(kn^2\kappa)$ & $O(kn^2\kappa)$\\
\textsf{Encode} & $O(m\kappa)$ & 0 \\
\textsf{Adding $n'$ users} & $O(n'n)$ & $O(n'n)$ \\
\hline\hline
\end{tabular}
\end{table}

User 1 with special role has to receive all other users' communication strings to generate the homomorphic ciphertexts, and his communication overhead is $\theta_{min}$ times greater. Besides, we only need two communication rounds for each polynomial evaluation, which is a constant irrelevant to the number of participants. Specifically, everyone except user 1 sends out his data in the first round, and user 1 broadcasts the homomorphic ciphertext in the second round. Finally, when $\kappa=512$, the actual size of an encoded value $C(x_{ik})$ is 256B, and a homomorphic ciphertext $E(C(x_{2k}))$ is 512B.

\subsection{Practical Performance: Computation}

To evaluate the computation overhead, we implemented our scheme in Ubuntu 12.04 using the GMP library (gmplib.org) based on C in a computer with Intel i3-2365M CPU @ 2.80 GHz $\times 2$, Memory 3GB. To exclude the communication overhead from the measurement, we generated all the communication strings as files in one computer and conducted all the computation locally (I/O overhead excluded in the measurement). Security parameter $\kappa$ is set as 512 bits to achieve strong prime numbers (the moduli $N,\tilde{N}$ are around 1024 bits). Paillier's cryptosystem is employed as the homomorphic encryption in \textsf{Encode}, which is also implemented in C.

\subsubsection*{Microbenchmark}

We first perform a series of microbenchmarks to measure the basic cost units of our scheme as in Table \ref{table:micro}. Overhead of the \textsf{Encode} and \textsf{Decode} algorithms depend on the polynomial to be evaluated, so we randomly chose a poylnomial with single product term having 10,000 participants to measure the exact run time of each algorithm (Table \ref{table:micro}).
\begin{table}[h]\small
\centering
\caption{Basic Cost Units}\label{table:micro}\vspace{-10pt}
\begin{tabular}{r|c|c|c|c}
\hline\hline
Algorithm & Min & Max & Mean & Std. Dev. \\\hline
\multicolumn{5}{c}{Authority}\\\hline
\textsf{Aggregate} & 0.25ms & 0.31ms & 0.28ms & 0.01ms \\\hline
\multicolumn{5}{c}{Special user 1}\\\hline
\textsf{Encode} & 9.7ms & 10.1ms & 9.846ms & 0.057ms\\\hline
\multicolumn{5}{c}{Special user 2}\\\hline
\textsf{Encode} & 9.4ms  & 9.6ms & 9.458ms & 0.053ms\\\hline
\multicolumn{5}{c}{Other participants}\\\hline
\textsf{Encode} & 0.115ms  & 0.145ms & 0.129ms  & 0.0186ms\\\hline\hline
\end{tabular}
\end{table}

User 1's and 2's overhead is more expensive than other users because they conduct operations in both $\mathbb{Z}_{N}^*$ (1024 bits) and $\mathbb{Z}_{N^2}^*$ (2048 bits) while other users' operations are closed only in $\mathbb{Z}_N^*$. Besides, the aggregator's overhead is cheaper than user 1's or 2's because he only performs multiplications which are much cheaper than exponentiations.

As discussed in Section \ref{section:minimum_threshold_optimization}, User 1's overhead is affected by the number of actual participants in each product. Also, it is clear that the overall overhead is related to the number of products in the polynomial.
To evaluate the scalability \textit{w.r.t.} those parameters, we execute the \textsf{Encode} algorithm and \textsf{Aggregate} algorithm for randomly chosen polynomials with different parameters (Figure 1).

\begin{figure*}[t]
  \centering
  \begin{tabular}{ccc}
  \includegraphics[width=0.23\textwidth]{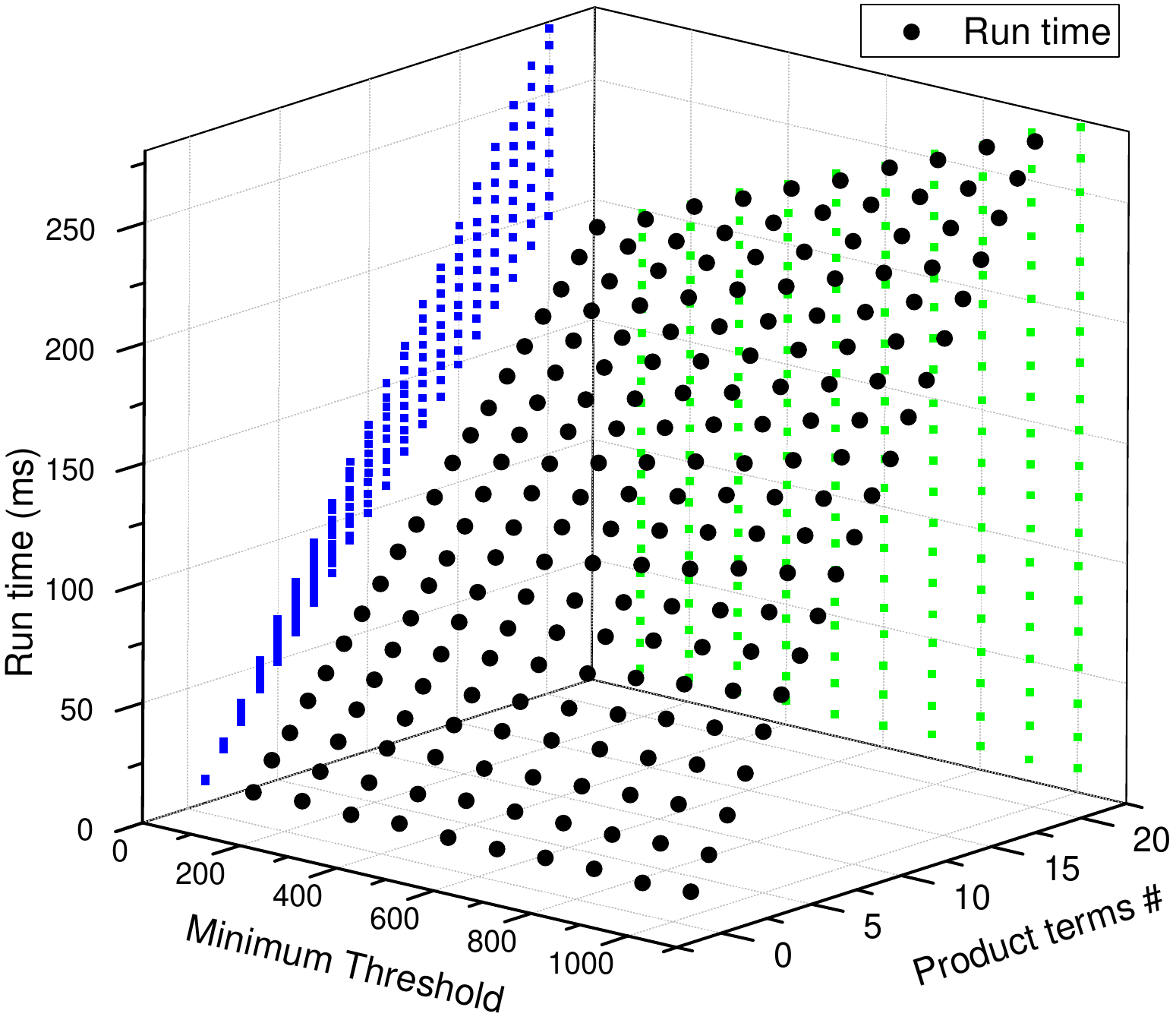} &
    \includegraphics[width=0.26\textwidth]{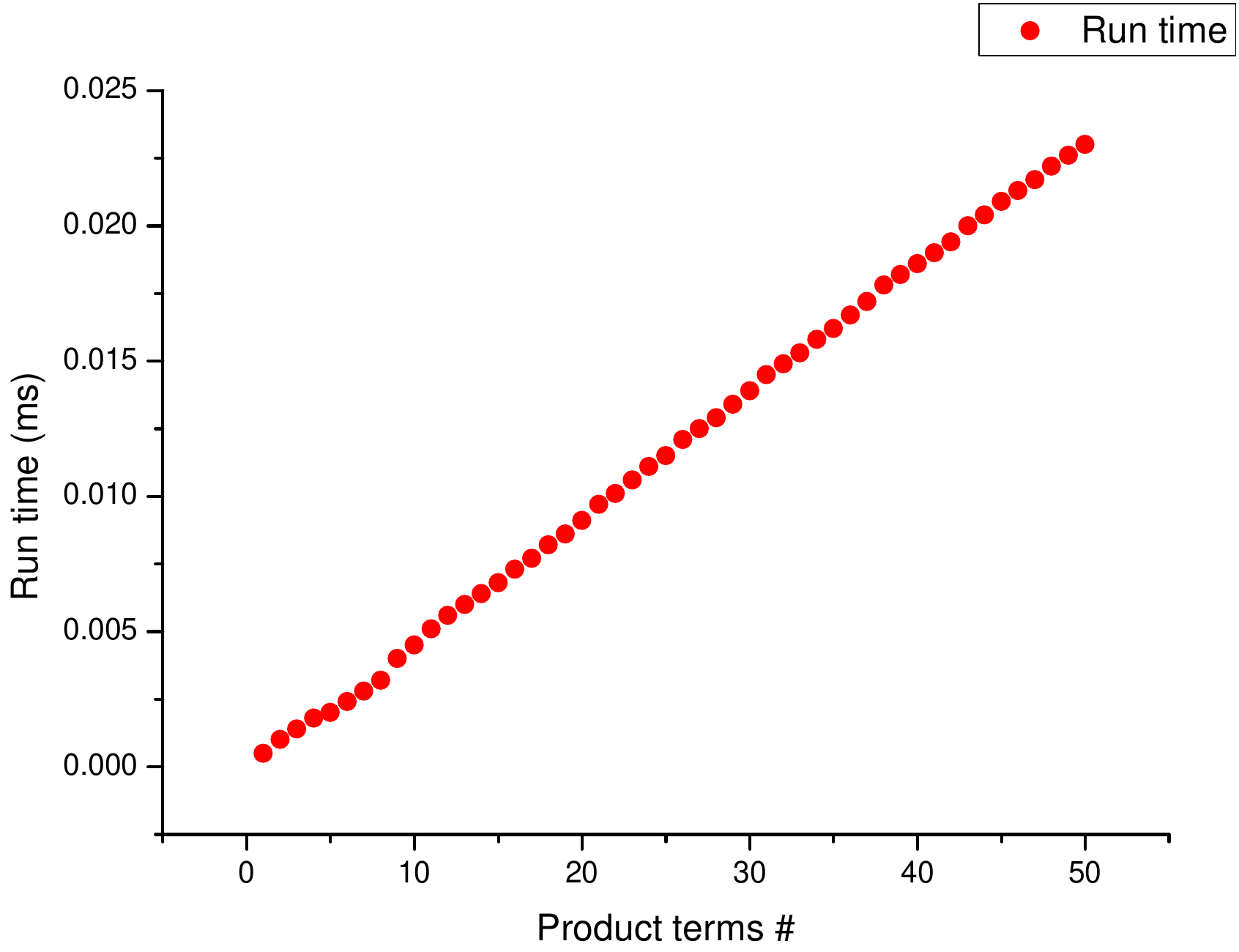} & \includegraphics[width=0.26\textwidth]{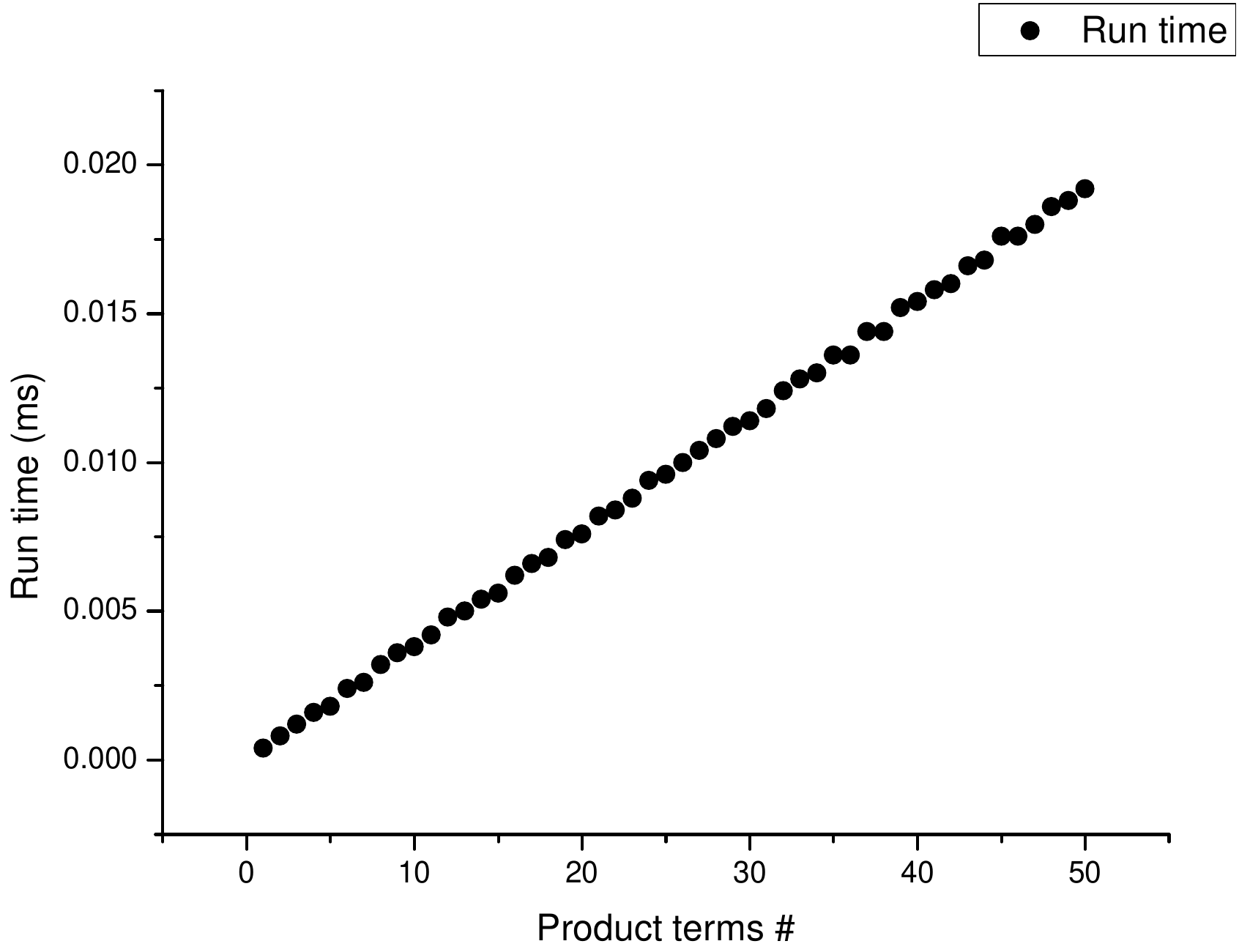} \\
(a) User 1's \textsf{Encode} algorithm & (b) Other users' \textsf{Encode} & (c) Aggregator's \textsf{Aggregation}
\end{tabular}\vspace{-5pt}
  \caption{Run time of different algorithms for different roles.}\vspace{-5pt}
  \label{fig:prod}
    \label{fig:sum}
\end{figure*}

In Figure 1(a), $x$-axis is the minimum threshold, $y$-axis is the number of product terms in a polynomial, and $z$-axis is the run time. In Figure 1(b) and 1(c), $x$-axis is the number of product terms in a polynomial, and $y$-axis is the run time. Clearly, the run time of \textsf{Encode} grows bilinearly \textit{w.r.t.} the parameters for the special-role user 1, therefore, when either parameter is large, one should employ a cloud server to act as the special user 1 since an ordinary user may not be able to handle such huge computation. The run time of \textsf{Encode} grows linearly \textit{w.r.t.} the number of product terms for other users (Figure 1(b)), and the same applies for \textsf{Aggreate} for the aggregator.   Note that the performance is more sensitive to the number of products because it is equal to the number of exponentiations in $\mathbb{Z}_{N^2}^*$, which is much more expensive than the multiplications.

\subsubsection*{Data Analysis via \ourprotocol}

In this section, we obtain various datasets available online (UCI machine learning datasets \cite{Bache+Lichman:2013}, SNAP Amazon Movie Review \cite{mcauley2013amateurs}, SNAP Amazon Product Review \cite{mcauley2013hidden}), and conduct various analysis on them as described in Section \ref{section:applications} to show the generality of our framework as well as its performance. Note that the run times shown below are the overall run time when all the computations are conducted in a laptop computer, and they are not the run times per user. The actual run time for non-special user is almost negligible because the special users' overhead contribute to the majority of the run time as analyzed above.

\noindent \textbf{Statistics Calculation}:  As aforementioned (Section \ref{section:applications}), various statistics analysis are represented as polynomials, which can be directly evaluated by our scheme.  To compare the performance with peer works, we implemented three schemes Shi \cite{shi2011privacy}, Joye \cite{joye2013scalable}, and Jung \cite{jung2014collusion} in the same environment, whose purpose is to calculate a product or a sum on a private dataset. Due to this limitation, these schemes can be used to calculate simple statistic values only (\textit{e.g.,} mean, standard deviation), and we compared the overall run time of the mean calculation on random values as most of the statistics analysis supported by \cite{joye2013scalable,jung2014collusion,shi2011privacy} are based on mean calculation. The entire calculation is conducted on a laptop computer to measure the pure computation overhead only, and the security parameter of each scheme is chosen such that all schemes have the similar security level ($\approx$80-bit security level).
\begin{figure}[h]
\centering
\includegraphics[scale=0.25]{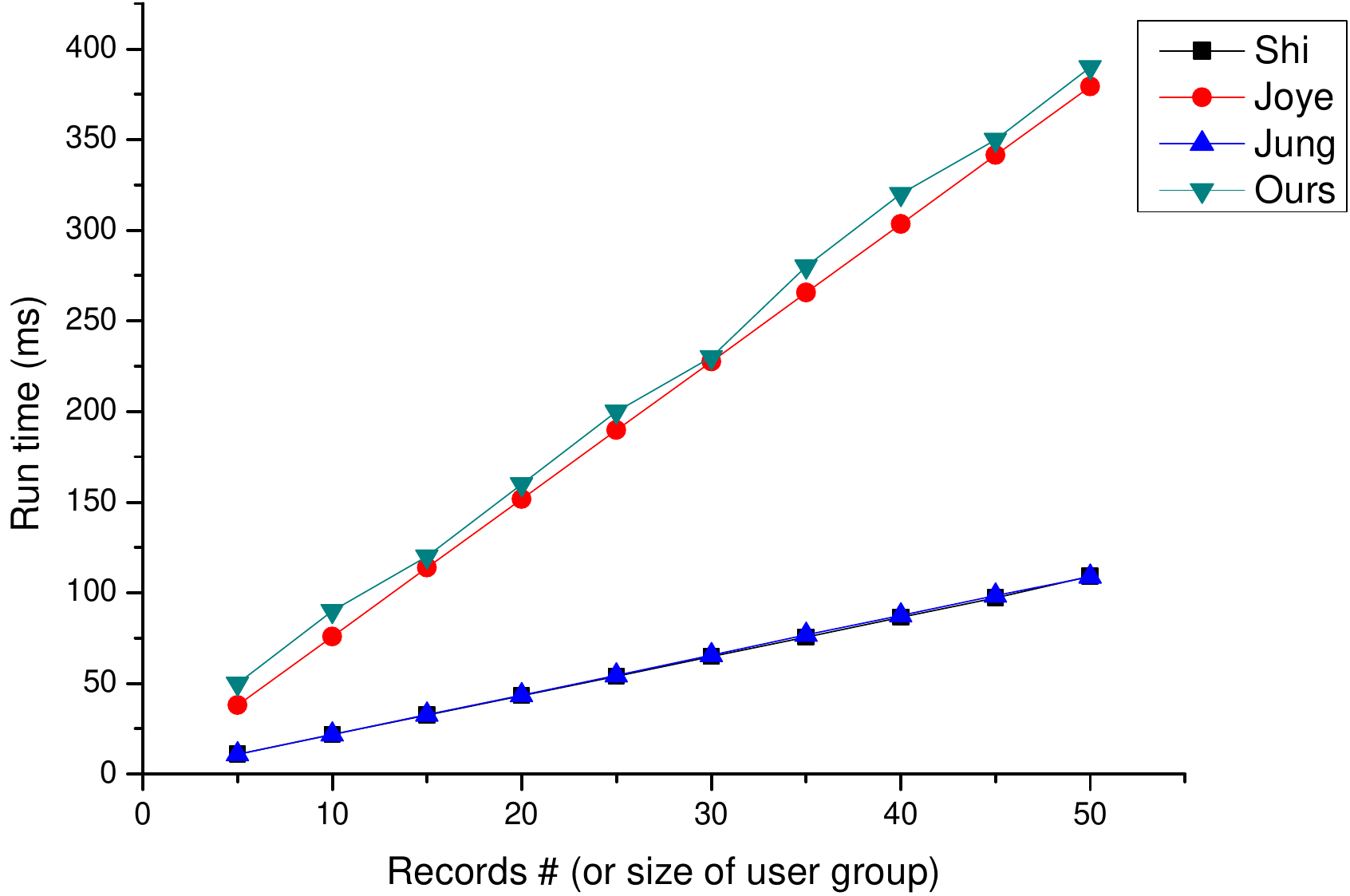}\vspace{-5pt}
\caption{Comparisons of mean calculation}\label{figure:mean_compare}\vspace{-5pt}
\end{figure}

As shown in Figure \ref{figure:mean_compare}, the overhead of our scheme is comparable to the one of Joye \cite{joye2013scalable}, and is twice of the ones of Shi \cite{shi2011privacy} and Jung \cite{jung2014collusion}. The main factor of this gap is the bit-length of the moduli. When bit-security levels are comparable to each other, the bit-lengths of the largest moduli in Shi \cite{shi2011privacy} and Jung \cite{jung2014collusion} are as twice large as the ones in Joye \cite{joye2013scalable}. Although our performance is not the best among these four schemes, other schemes can only let an aggregator obliviously evaluate the product or the sum over a fixed dataset. Considering our scheme's generality which enables an aggregator obliviously evaluate general multivariate polynomials over arbitrary subset of datasets (with the support of new user arrival and old user removal), we claim that this performance gap is acceptable.

\noindent \textbf{Regression Analysis}: We trained a linear regression model over the datasets in a privacy-preserving manner using our scheme (Section \ref{section:applications}), and measured the communication overhead for each ordinary user as well as the overall run time of the entire training in a local computer (Table \ref{table:lr_data}). Specifically, in SNAP datasets, we counted the number of `good' strings (\textit{e.g.,} `I love', `I like', `best') and `bad' strings (\textit{e.g.,} `I hate', `terrible') in the review, and treated those numbers as well as the price as the features. We further treated the score rating as the dependent variable.

\begin{table}[h]\small
\centering\caption{Linear Regression on Datasets}\label{table:lr_data}\vspace{-10pt}
\begin{tabular}{c|c|c|c|c}\hline\hline
Datasets & Records \# & Feature \# & Comm. &  Time \\\hline
\multicolumn{5}{c}{UCI Dataset}\\\hline
Adult & 48,842 & 14 & 237KB &  355s \\
Bank & 45,211 & 17 & 344KB & 341s\\
Insurance & 9,822 & 14 & 236KB &  74s \\
White wine & 4,898 & 11  & 148KB & 33s \\
Red wine & 1,599 & 11 & 148KB & 12s \\\hline
\multicolumn{5}{c}{SNAP Dataset}\\\hline
Arts & 28K & 3 & 13KB & 210s \\
Books & 13M & 3  & 13KB & 26h \\
Games & 463K & 3 & 13KB & 0.9h\\
Movie & 8M  & 3  & 13KB & 11h \\
\hline\hline
\end{tabular}
\end{table}

Moreover, we compare our performance with a similar work \cite{nikolaenko2013privacy} which designed a privacy-preserving ridge regression over millions of private data records (Table \ref{table:linear_compare}). As we failed to obtain the source codes of their scheme, we perform the comparison based on the statistics provided in their paper. This is not a completely fair comparison of the different computing environment. However, \ourprotocol~supports any polynomial-based data analysis while \cite{nikolaenko2013privacy} only supports the ridge regression, therefore we claim that our solution is more preferable if the performance of our general solution is comparable to the ad hoc solution.

\begin{table}[h]\small
\centering
\caption{Comparison with \cite{nikolaenko2013privacy}}\label{table:linear_compare}\vspace{-10pt}
\begin{tabular}{c|c|c|c}\hline\hline
Ridge Regression & Insurance & Red wine & White wine\\\hline
\cite{nikolaenko2013privacy} & 55s & 39s & 45s\\\hline
Ours & 74s & 12s & 33s \\\hline\hline
\end{tabular}
\end{table}

\noindent \textbf{SVM Analysis}: We implemented the SMO algorithm \cite{platt1998sequential} in a privacy preserving manner (Section \ref{section:applications}) to train a linear support vector machine over the same datasets (part of UCI datasets) as those in \cite{platt1998sequential}. Please refer to the paper for the detailed description of the dataset.

\begin{table}[h]\small
\centering\caption{SMO on Datasets}\label{table:smo_data}\vspace{-10pt}
\begin{tabular}{c|c|c|c}\hline\hline
Datasets & Records \# & Features \# &  Time \\\hline\hline
Adult \cite{kohavi1996scaling}  & 1,605 & 14 &  2.75h \\
Web \cite{breese1998empirical}  & 2,477 & 17 &  49.75h\\
\hline\hline
\end{tabular}
\end{table}

Note that the run time to train a SVM is linear to the number of iterations until convergence in the SMO algorithm, which heavily depends on the characteristics of the dataset. Our experiment shows that the SVM training via \ourprotocol~has a non-negligible overhead (Table \ref{table:smo_data}), but this is inevitable because 1) the run time of plain SMO algorithm on the same dataset is also of several seconds, and 2) the inherent complexity of the multiplications and exponentiations over $k$-bit integers are $O(\kappa^2)$ and $O(\kappa^3)$ respectively, and the plain SVM without privacy consideration performs arithmetic operations in double type numbers (8B) while our framework (with 80-bit security level) performs arithmetic operations as well as exponentiations on 256B numbers. To the best of our knowledge, \cite{laur2006cryptographically} is the first and only work who presents a privacy-preserving method to train a SVM over a large number of individual records (most of similar works study how to train SVM on a small number of vertically or horizontally partitioned database), but their work does not evaluate the performance in an real implementation. Although we were not able to acquire the implementation, we could indirectly compare our performance with this only similar work. Their theoretic performance complexities (computation \& communication) in theory are exponentially greater than ours. Such a gap exists because they relied on the secure multi-party computation. The performance gap between the solutions based on secure multi-party computation and our work discussed in Section \ref{sec:smc_gap} also tells the performance gap in real implementations is huge. Therefore, ours can be considered as a development to their solution which improved the performance aspect by getting rid of the secure multi-party computation.

\section{Discussions \& Analysis}\label{section:discussion}

\subsection{Disjoint Time Domains}\label{section:disjoint_time}

To be indistinguishable against CPA, our \textsf{Encode} must be randomized. However, it is not easy to let users achieve independent but coordinated random masks such that their aggregation yields a constant if we do not rely on a trusted center. We use the time slot as the `seed' of our hash function $H(\cdot)$ to address this challenge. 

In the $\mathsf{Encode}$ algorithm, $H(t_k)^{q^{(|\mathcal{P}|-1)}(i)\mathcal{L}_{i,\mathcal{P}}}\in\mathbb{Z}_N^*$ is used to mask the private value $x_{ik}$. As analyzed in the appendix, such random masks are seemingly random under the random oracle model. Since the random mask is based on the hash function $H(\cdot)$ which is deterministic, we require a disjoint time domain and forbid the re-use of any time slot $t_k$. Same time slots yield same randomizers, and this will allow some polynomial adversaries to have non-negligible advantages in our game.

\subsection{Minimum Threshold for $|\mathcal{P}|$}\label{section:minimum_threshold}

Since the security of \ourprotocol~relies on the random masks whose product or sum is equal to an identity element, we require the minimum number of users in any polynomial evaluation should be $\theta_{min}=3$. Otherwise, a user can trivially deduce another user's secret random mask $r$ by inverting his own random mask $r^{-1}$ (in multiplicative group). That is, besides the special user 1\&2, we also need one ordinary user participating in the evaluation. This is why we calculate the secret key $q^{(d)}(i)$ in $\mathsf{KeyGen}$ algorithm only for $d=2,3,\cdots,n-1$ (recall that $q^{(k)}(i)$ is used when $k+1$ users participate in the evaluation).

%\subsection{User Revocation}
%User revocation \cite{ren2008efficient} usually incurs extra burden to the key management system because a revoked user $\mathcal{R}$'s key should be unusable any longer . In \ourprotocol, every participant $i$'s random noise $H(t)^{q^{(|\mathcal{P}|-1)}(i)\mathcal{L}_{i,\mathcal{P}}}$ is calculated by $i$ based on the $\mathcal{P}$, where $H(\cdot)$ is employed as the random oracle. Therefore, as long as other participants do not include the revoked user $\mathcal{R}$'s ID in $\mathcal{L}_{i,\mathcal{P}}$, $\mathcal{R}$ can neither infer others private input or illegally inject his data into the analysis.

\subsection{Passive Rushing \& Collusion Attacks}\label{section:attacks}

In previous sections, we intentionally simplified the adversarial model as well as the construction for the readability. Consequently, the basic construction allows certain attacks. In this section, we present a countermeasure to achieve a more complete construction.

In a passive rushing attack, attackers access the messages sent from others before they choose their random numbers \cite{feigenbaum1991distributed}. Specifically, during the exchanges of $Y_j$ in the $\mathsf{KeyGen}$ algorithm, an adversarial user $i-1$ can send out $y_{i-1}=y_{i+1}^{-a}$ to user ${i}$ ($a\in\mathbb{Z}_{\tilde{N}}^*$) after receiving $y_{i+1}$. Then, $\participant_{i}$'s random mask is equal to $y_{i+1}y_{i-1}^{-1}=\tilde{g}^{(r_{i+1}-(r_{i+1}-a))r_{i}}=\tilde{g}^{ar_{i}}$, which can be efficiently calculated with $i$'s public parameter $y_i=\tilde{g}^{r_i}$. This attack does not tamper the final computation and cannot be detected. With exactly same theory, two colluding users $i+1$ and $i-1$ can also easily calculate $i$'s secret parameter $Y_i=\tilde{g}^{(r_{i+1}-r_{i-1})r_i}$.

However, we can adopt the ideas from \cite{jung2014collusion}. With one extra round of exchange, we can let each user $i$ have $Y_i=(y_{i+2}y_{i-1}^{-1})^{r_{i+1}r_i}$, and a passive rushing attacker or two colluding users will fail to infer a user's $Y_i$. In general, to defend against $k$ colluding passive rushing attackers or $k+1$ colluding users, we can have $k$ more extra rounds of exchanges to let each user $i$ have
$Y_i=\left(y_{r_{i+k+1}}y_{r_{i-1}}^{-1}\right)^{r_{i+k}r_{i+k-1}\cdots r_{i+2}r_{i+1}r_i}$.

Recall that we initially assumed non-cooperative users in this paper. But with the above prevention, we can relax the assumption and make our \ourprotocol~resilient to a collusion attack of up to $k-1$ passive rushing attackers or $k$ normal attackers. As a result, all encoding keys $q^{(2)}(i),\cdots,q^{(k)}(i)$ become unusable. This is because any user's encoding key $\SK_i$ is composed of several data points in hidden polynomials, and $k$ colluding users can deduce the $k$-degree hidden polynomial $q^{(k)}(x)$ of 0 constant term. Then, these colluding users can compute any user $i$'s encoding key $q^{(k)}(i)$. Thus, the lowest degree of the usable polynomial is $k+1$, and the minimum threshold $\theta_{min}$ for $|\mathcal{P}|$ in the \ourprotocol~becomes $k+2$ (each user uses $q^{(k+1)}(i)$ as the encoding key) to guarantee the semantic security of \ourprotocol. Employment of this countermeasure translates to multiple extra communication rounds in \textsf{KeyGen}, but this is a one-time operation at the beginning of the system initialization.

\subsection{Performance Optimization}\label{section:minimum_threshold_optimization}

In the algorithm \textsf{Encode}, we require that everyone should submit his encoded value $C(x_{ik})$ even though his $x_{ik}$ does not appear in the product term $\prod x_{ik}^{e_{ik}}$ (in which case $x_{ik}$ is set as 1). This is because the number of actual participating users should be at least a certain threshold $\theta_{min}$ to guarantee the semantic security (reviewed in details in Section \ref{section:limit}). However, in some polynomials, single product terms involve very few users regardless of the total number of participants $|\mathcal{P}|$ of the polynomial, such as
$f(\mathbf{x}_\mathcal{P})=\sum_{i,j\in\mathcal{P}} x_ix_j$. If strictly following \textsf{Encode}, one needs to have $|\mathcal{P}|-1$ (encoded) 1's multiplied for each product term $x_ix_j$. However, to preserve the semantic security of the scheme, we only need $\theta_{min}$ users to participate in each single product term, and this is an unnecessary overhead. This performance degradation becomes huge when $\theta_{min}\ll |\mathcal{P}|$, but we can simply get rid of it as follows. Whenever the number of participants involved in a product term is less than $\theta_{min}$, randomly find dummy users in $\mathcal{P}$ only until we have $\theta_{min}$ users instead of having everyone in $\mathcal{P}$ participating in the analysis. Then, dummy users participate in the evaluation by providing 1 as their private values, and by doing so, we only have minimum number of participants for every product term, which greatly cut off the unnecessary overhead.

\subsection{Limitations \& Future Works}\label{section:limit}
\noindent \textbf{Performance Limit}:~It is shown in Section \ref{sec:evaluation} that user 1's communication and computation overhead grows linearly \textit{w.r.t.} the actual participants in each product term, which may become huge in some big data analysis. In such case, we may add a cloud server to take this role, and let him provide 1 during the evaluation. Besides, each person needs to send/receive $O(n^2\kappa)$ bits (Section \ref{sec:theoretic_analysis}) with small constant during the one-time operation \textsf{KeyGen}, which may be large when $n$ is large. This can be reduced with several na\"{i}ve ideas, for examples, dividing $n$ users into several subgroups and forcing the analysis within the subgroups only. But such ideas disable some types of analysis, and handling this limit is one of our future works.\\
\textbf{Polynomial Parameters}:~This paper focuses more on user's data privacy, and pays less attention to the aggregator's privacy. This is why we require that the polynomials should be public, otherwise aggregators may adaptively design the polynomial (\textit{e.g.,} $f(\mathbf{x}_{\mathcal{P}})=x_{1}$) to directly infer a user's data. However, in some cases, the polynomial parameters may be proprietary property which also need to be protected (\textit{e.g.,} data mining techniques), but then corresponding verification mechanisms should be introduced as well to enable users to verify whether the polynomial is maliciously designed, which is one of our future works.\\
\textbf{Finite Real  Numbers}: Although our design supports polynomials over real numbers, the real numbers are represented via fixed-point representation, which constructs one-to-one mapping between the real numbers and the integers in the cyclic group we use as a part of crypto protocols. Therefore, the real numbers that can be expressed in our system is a finite set of real numbers whose size is equal to the order of the group in our algorithms (\textit{i.e.,} $N$).\\
\textbf{Data Publishing}:~Our \ourprotocol~only supports  \textit{pull-based} data analysis in which the aggregator requests the data analysis with the declaration of the polynomial, and then users publish their data according to it. One thought is to achieve a \textit{push-based} analysis, where users publish their data whenever it is generated, and the aggregator chooses desired subset of data and periods to perform the data analysis. However, once the data is published, it will not be under the user's control, and it becomes much more challenging to protect the privacy. Designing a push-based data analysis which can prevent malicious analysis on the already published data is a very challenging task, and we leave this as our future work too.

\begin{table*}[t]\small
\begin{center}
\caption{Comparisons}\label{table:comparison}\vspace{-10pt}
\begin{tabular}{r|c|c|c|c}
\hline\hline
Approach based on & Storage Complexity & Communication Rounds & Accurate result? & Rely on secure channel? \\\hline
Multi-party Computation & High & $O(n)^{*}$ & Yes & No \\
Perturbation & $O(1)$ & $O(1)$ & No & No \\
%Segmentation & 0  & $O(1)$ & Yes & Yes\\
%Homomorphic Encryption & $O(1)$ & $O(n)$ & Yes & Yes\\
Functional Encryption & High & O(1) & Yes & Yes\\
Secret-shared Keys & $O(2^n)$ & $O(1)$ & Yes & Yes, except \cite{jung2014collusion,jung2013privacypreserving}\\
\textbf{Ours} & $O(n)$ & $O(1)$ & Yes & No\\\hline\hline
\end{tabular}
\end{center}
*This is the best known complexity, and every gate in the circuit incurs one round of oblivious transfer.
\end{table*}

\section{Related Work}\label{section:related}

The purpose of our scheme is to protect the data privacy rather than the identity privacy. Therefore, we focus more on reviewing peer works studying the privacy-preserving data analysis whose purpose is to protect the data confidentiality. In general, related approaches can be categorized into the following families: approaches based on secure multi-party computation, perturbation, segmentation,  functional encryption and secret-shared keys.

\subsection{Secure Multi-Party Computation}\label{sec:smc_gap}

\begin{table}[h]\centering\caption{Comparison of the fastest SMC implementation and ours}\label{tab:comparison-smc}\vspace{-10pt}
\begin{tabular}{c|c}
\hline\hline
5,550 AND via GMW \cite{choi2012secure} & 134 additions via ours\\\hline
15-20 seconds & 0.06-0.08 seconds \\\hline\hline
\end{tabular}
\end{table}

Secure Multi-party Computation  (SMC) \cite{yao1982protocols}  enables $n$ parties to jointly and privately compute a function $f_i(\mathbf{x})$ where the result $f_i$ is returned to the participant $i$ only. One can trivially use SMC to enable data analysis on published sanitized data, but SMC is subject to extraordinarily high delay in real life implementations. Among various real implementations on multi-party computation schemes (FairplayMP \cite{ben2008fairplaymp}, VIFF \cite{damgaard2009asynchronous}, SEPIA \cite{burkhart2010sepia}, and GMW \cite{choi2012secure}), even the fastest GMW \cite{choi2012secure} does not scale well, whose communication rounds is linear to the number of users with huge constant factors, and it takes 15-20 seconds to conduct 5,500 AND gates evaluation. Similar-size computation in our scheme only incurs subsecond-level computation time and single-round communication of several milliseconds (Table \ref{tab:comparison-smc}). Such huge performance gap is due to the different purposes. SMC targets at implementing arbitrary computation while ours target at polynomial-based data analysis only. Considering the number of users and the volume of data, it is more practical to rely on schemes specially designed for general analysis to enable lightweight protocols.

\subsection{Perturbation}

The perturbation is often leveraged to achieve the differential or membership privacy (\cite{dwork2008differential,li2013membership}). In such approaches, a random noise drawn from a known distribution is added to mask the original data, and some statistics about the aggregate result can be deduced based on the distribution of the random noises. These approaches enable certain analysis on published (sanitized) data (normally without requiring a trusted key dealer), but the accuracy is dependent on the guaranteed security (\textit{i.e.,} accuracy increases when security is sacrificed), and the temporal correlation between the data leads to a limited utility when the data is time-series. \'{A}cs \textit{et al} proposed a privacy-preserving smart metering system which support various aggregated statistics analysis without leaking household activities. \cite{acs2011have} presented how to perform Fan \textit{et al.} \cite{fan2013adaptive} presented an adaptive approach to time-series data releasing with differential privacy, but their relative error is greater than $10^{-1}$ with certain privacy sacrifice. Recently, Chen \textit{et al.} \cite{chen2012differentially} presented a scheme based on perturbation to mine frequent patterns in the form of substring, but the utility for prefixes is poor due to the noises making the scheme differentially private. Supported by the recent demand in the accuracy \cite{ioannidis2014privacy}, we claim that, for some applications in the future, it is not desirable to rely on the approaches whose accuracy depends on the privacy requirements of the applications.

\subsection{Functional Encryption}\label{sec:FE}

Functional encryption \cite{boneh2011functional} is a type of recently proposed public key encryption in which a certain function of the plaintext can be derived from the ciphertext if one possesses a valid private key corresponding to the function. It is possible to leverage FE to design a solution to our problem, but the one key is associated with a pre-defined function in FE, and this key can be used to derive only one function of the plaintext. In many data mining algorithms, the actual target analysis functions (classifiers or predictive models) are trained iteratively, in which each training at each iteration is another type of function to be evaluated over the private dataset. Since all such analysis are data-dependent, issued keys can hardly be re-used, and keys should be issued to all participants in advance of every data analysis, which is a communicationally expensive interaction. Furthermore, the key distribution must be performed via secured channels.

In contrast, our keys can be re-used arbitrarily many times as long as the time windows for every polynomial are disjoint with each other.

\subsection{Secret-shared Keys}

Several schemes distribute certain keys to the data users such that the sum or product of the keys are constants. Then, these keys are used to mask the original data. These schemes are very close to ours.

Shi \cite{shi2011privacy} and Joye \cite{joye2013scalable} let the aggregator compute the sum of $n$ participants' data $x_i$'s without learning anything else. They both let a trusted third party generate random keys $s_i$ for the $i$-th participant such that $\sum s_i=0$. Then, it follows that $\prod H(t)^{s_i}=1$, where $H(\cdot)$ is a hash function employed as a random oracle and $t$ is the time slot when the calculation is carried out. Then, this property is used by both schemes to compute the sum $\sum x_i$, and both of them are proved to be semantically secure against chosen-plaintext attack. However, both of the schemes rely on a trusted key issuer, and the data is not confidential against him. Furthermore, only the product or sum calculation is enabled, which makes it less suitable for a general data analysis.
Jung \cite{jung2014collusion,jung2013privacypreserving}
removed the necessity of a trusted key dealer in their settings, and their scheme enables both product and sum calculations. They also mention the possibility of evaluating some multivariate polynomials, but all the product terms in the polynomials are revealed to the aggregator. More importantly, the hardness of breaking their scheme reduces to computational diffie-hellman (CDH) problem only, which makes their scheme one-way but not semantically secure.
%In our scheme, we employ the multivariate polynomial evaluation protocol \cite{jung2013privacypreserving} as a building block to achieve the accurate data aggregation within constant communication rounds. The trusted third party in \cite{shi2011privacy} is removed since data privacy against the aggregator is also a top concern, and unlike \cite{he2007pda}\cite{sheikh2009privacy}, we assume insecure channels, which enabled us to get rid of  expensive and vulnerable key pre-distribution. We did not segment each individual's data, our protocols only incur constant  communication overhead for each participant. Our scheme is also based on the hardness of the discrete logarithm problem like \cite{shi2011privacy}, but we do not trivially employ brute-force manner in decryption, instead, we employ our novel efficient protocols for sum and product calculation. 

These four schemes based on secret-shared keys are close to our solution, but all the schemes need a fixed group and assign a set of secret keys, which means for a set of $n$ users, one needs to let all users hold $\theta(2^n)$ keys to conduct data analysis among any subgroup of them, which violates the flexibility requirement.\smallskip

%Above two schemes are both lack of flexibility in a sense that it is intractable to conduct data aggregation among any combination of the participants. In order to support aggregation among any subset of $n$ participants, each participant in both schemes needs to store $2^n$ set of keys for all possible combinations $2^{\{1,\cdots,n\}}$ of the participants.

As a summary, the comparisons of all aforementioned approaches and ours are shown in the Table 6.

\subsection{Secure Multivariate Polynomial Evaluation}
To the best of our knowledge, \cite{blanton2011achieving,dachman2011secure,franklin2010efficient,jung2013privacypreserving} are the only four works who study the secure multi-party computation on multivariate polynomial evaluation. First of all, the solution in \cite{jung2013privacypreserving} reveals all single product terms in the polynomial, and it leaks much information about the private values. \cite{franklin2010efficient} focused on a polynomial of degree 3, and points out that this can be generalized to a higher degree. However, in higher degrees, the communication overhead is not optimal. Subsequently, \cite{dachman2011secure} presented a more efficient scheme with the unicast complexity  $O(\kappa m^2n\log d)$ and broadcast complexity $O(\kappa m^3(\log d)^2)$ (untight bound for simplicity), where $\kappa$ is the security parameter, $m$ is the number of product terms in the polynomial, $n$ is the number of participants, and $d$ is the highest degree of the polynomial. Also, their communication rounds is a constant in terms of the `round table' rounds, which is linear to the number of users. In our framework, the unicast complexity for each evaluation is only $O(\theta_{min}m\kappa)$ for the special user ($\theta_{min}$ being the minimum threshold for the number of participants) and $O(m\kappa)$ for other participants, and there is no broadcast overhead. Also, the communication round is a constant irrelevant to the number of participants in the evaluation. Besides, \cite{blanton2011achieving} also discusses how to compute limited families of polynomials, but the rounds complexity is $O(l\log l)$ where $l$ is the bit-length of the inputs.

\section{Conclusion}\label{section:conclusion}

In this paper, we successfully designed \ourprotocol, a general framework which enables a third-party aggregator to perform many types of polynomial-based data analysis on private data records, where the data records are possessed by millions of users. Namely, our framework enables the aggregator to evaluate  various multivariate polynomials over any subgroup of users' private data without learning  individuals' input values, and it also supports new users to join the system and old users to leave the system. Our formal proof shows the semantic security of \ourprotocol~against chosen-plaintext attacks, and the extensive evaluations show that \ourprotocol~incurs constant overhead regardless of the total number of users or the complexity of the polynomial. Besides, \ourprotocol~also exhibits various nice properties (design goals in Section \ref{section:introduction}) which make it suitable for real life applications.\vspace{-5pt}

{\small
\bibliographystyle{abbrv}
\bibliography{CombinatorialAggregation}
}
\vspace{-0.3in}

% Generated by IEEEtranS.bst, version: 1.13 (2008/09/30)

\begin{IEEEbiography}[{\includegraphics[width=1in,height=1.25in,clip,keepaspectratio]{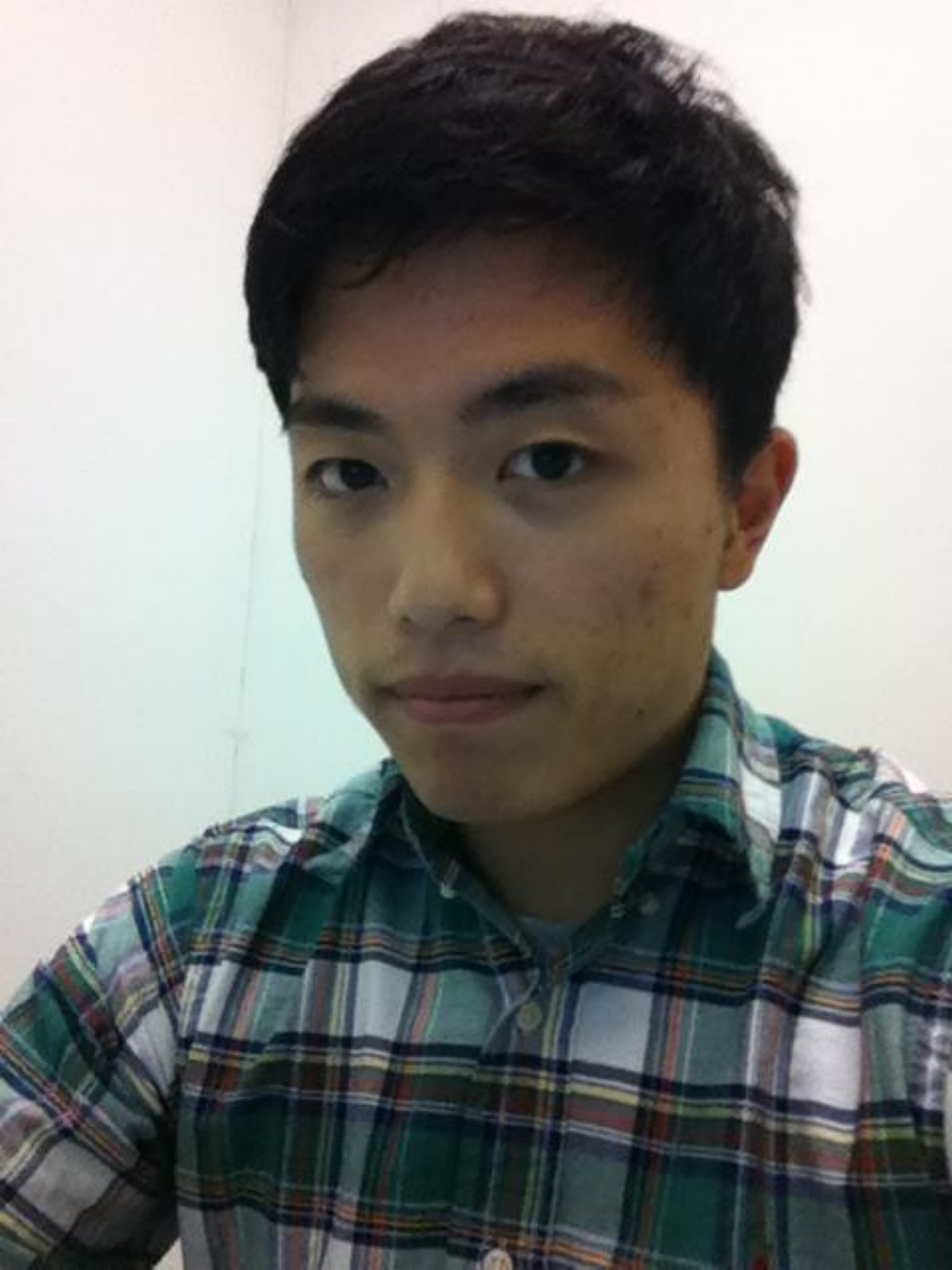}}]{Taeho Jung}
Taeho Jung received the B.E degree in Computer Software from Tsinghua University, Beijing, in 2011, and he is working toward the Ph.D degree in Computer Science at Illinois Institute of Technology while supervised by Dr. Xiang-Yang Li. His research area, in general, includes privacy \& security issues in big data analysis and networking applications. Specifically, he is currently working on the privacy-preserving computation in various applications and scenarios where big data is involved.
\end{IEEEbiography}
\vspace{-0.5in}

\begin{IEEEbiography}[{\includegraphics[width=1in,height=1.25in,clip]{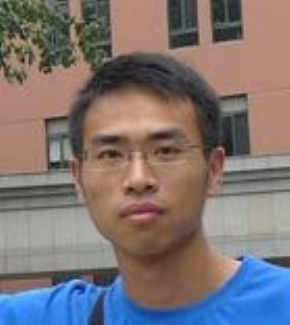}}]{Junze Han}
Junze Han is a Ph.D. student in Computer Science at Illinois Institute of Technology since 2011. He received the B.E. degree from the Department of Computer Science at Nankai University, Tianjin, in 2011. His research interests include data privacy, mobile computing and wireless networking.
\end{IEEEbiography}
\vspace{-0.5in}

\begin{IEEEbiography}[{\includegraphics[width=1in,height=1.25in,clip,keepaspectratio]{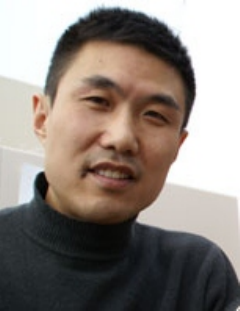}}]{Xiangyang Li}
Dr. Xiang-Yang Li is a professor at the Illinois Institute of Technology. He holds EMC -Endowed Visiting Chair Professorship at Tsinghua University. He currently is a distinguished visiting professor at Xi’An JiaoTong University, and University of Science and Technology of China. He is a recipient of China NSF Outstanding Overseas Young Researcher (B). Dr. Li received MS (2000) and PhD (2001) degree at Department of Computer Science from University of Illinois at Urbana-Champaign, a Bachelor degree at Department of Computer Science and a Bachelor degree at Department of Business Management from Tsinghua University, P.R. China, both in 1995. His research interests include wireless networking, mobile computing, security and privacy, cyber physical systems, and algorithms. He and his students won three best paper awards (ACM MobiCom 2014, COCOON 2001, IEEE HICSS 2001), one best demo award (ACM MobiCom 2012)and was selected as best paper candidates twice (ACM MobiCom 2008, ACM MobiCom 2005). He published a monograph "Wireless Ad Hoc and Sensor Networks: Theory and Applications". He co-edited several books, including, "Encyclopedia of Algorithms". Dr. Li is an editor of several journals, including IEEE Transaction on Mobile Computing. He has served many international conferences in various capacities, including ACM MobiCom, ACM MobiHoc, IEEE MASS. He is an IEEE Fellow and an ACM Distinguished Scientist.
\end{IEEEbiography}

\end{document}